\pdfoutput=1
\RequirePackage{ifpdf}
\documentclass{JINST}
\usepackage{caption}
\usepackage{booktabs}
\usepackage{lineno}
\usepackage{multirow}
\usepackage{url}
\usepackage{afterpage}

\title{Shower development of particles with momenta from 1 to 10 GeV in the CALICE Scintillator-Tungsten HCAL}

\author{\centering 
\LARGE\bf The CALICE Collaboration
}

\author{\centering
C.\,Adloff, 
J.-J.\,Blaising, 
M.\,Chefdeville, 
C.\,Drancourt,
R.\,Gaglione, 
N.\,Geffroy, 
Y.\,Karyotakis, 
I.\,Koletsou, 
J.\,Prast,
G.\,Vouters 
\\ \it
Laboratoire d'Annecy-le-Vieux de Physique des Particules, Universit\'{e} de Savoie,
CNRS/IN2P3,
9 Chemin de Bellevue BP110, F-74941 Annecy-le-Vieux CEDEX, France
}

\author{\centering
J.\,Repond, 
J.\,Schlereth, 
J.\,Smith$^a$, 
L.\,Xia 
\\ \it
Argonne National Laboratory,
9700 S.\ Cass Avenue,
Argonne, IL 60439-4815,
USA}

\author{\centering
E.\,Baldolemar, 
J.\,Li$^b$, 
S.\,T.\,Park, 
M.\,Sosebee, 
A.\,P.\,White, 
J.\,Yu 
\\ \it
Department of Physics, SH108, University of Texas, Arlington, TX 76019, USA
}

\author{\centering
G.\,Eigen 
\\ \it
University of Bergen, Inst.\, of Physics, Allegaten 55, N-5007 Bergen, Norway
}

\author{\centering 
M.\,A.\,Thomson, 
D.\,R.\,Ward
 \\ \it
University of Cambridge, Cavendish Laboratory, J J Thomson Avenue, CB3 0HE, UK
}

\author{\centering 
D.\,Benchekroun, 
A.\,Hoummada, 
Y.\,Khoulaki
 \\ \it
Universit\'{e} Hassan II A\"{\i}n Chock, Facult\'{e} des sciences, B.P. 5366 Maarif, Casablanca, Morocco
}

\author{\centering 
J.\,Apostolakis, 
D.\,Dannheim, 
A.\,Dotti, 
K.\,Elsener,
G.\,Folger, 
C.\,Grefe,
V.\,Ivantchenko, 
M.\,Killenberg$^c$, 
W.\,Klempt, 
E.\,van der Kraaij$^d$, 
C.B.\,Lam,
L.\,Linssen,
A.\,-I.\,Lucaci-Timoce$^e$, 
A.\,M\"{u}nnich$^c$, 
S.\,Poss,
A.\,Ribon,
A.\,Sailer, 
D.\,Schlatter, 
J.\,Strube,
V.\,Uzhinskiy
 \\ \it 
CERN, 1211 Gen\`{e}ve 23, Switzerland
}

\author{\centering
C.\,C\^{a}rloganu, 
P.\,Gay, 
S.\,Manen, 
L.\,Royer
 \\ \it
Clermont Universit\'e, Universit\'e Blaise Pascal, CNRS/IN2P3, LPC, BP
10448, F-63000 Clermont-Ferrand, France
}

\author{\centering
M.\,Tytgat,
N.\,Zaganidis
 \\ \it
Ghent University, Department of Physics and Astronomy,
Proeftuinstraat 86, B-9000 Gent, Belgium
}

\author{\centering
G.\,C.\,Blazey,
A.\,Dyshkant, 
J.\,G.\,R.\,Lima, 
V.\,Zutshi
 \\ \it
NICADD, Northern  Illinois University,
Department of Physics,
DeKalb, IL 60115,
USA
}

\author{\centering 
J.\,-Y.\,Hostachy, 
L.\,Morin
 \\ \it
Laboratoire de Physique Subatomique et de Cosmologie - Universit\'{e} Joseph Fourier Grenoble 1 -
CNRS/IN2P3 - Institut Polytechnique de Grenoble,
53, rue des Martyrs,
38026 Grenoble CEDEX, France
}

\author{\centering 
U.\,Cornett, 
D.\,David, 
A.\,Ebrahimi, 
G.\,Falley, 
N.\,Feege$^f$, 
K.\,Gadow, 
P.\,G\"{o}ttlicher, 
C.\,G\"{u}nter,
O.\,Hartbrich, 
B.\,Hermberg, 
S.\,Karstensen, 
F.\,Krivan,
K.\,Kr\"uger, 
S.\,Lu$^g$, 
B.\,Lutz,
S.\,Morozov, 
V.\,Morgunov$^h$, 
C.\,Neub\"user,
M.\,Reinecke, 
F.\,Sefkow, 
P.\,Smirnov,
M.\,Terwort
 \\ \it
DESY, Notkestrasse 85,
D-22603 Hamburg, Germany
}

\author{\centering  
E.\,Garutti, 
S.\,Laurien, 
I.\,Marchesini$^i$, 
M.\,Matysek, 
M.\,Ramilli
 \\ \it
Univ. Hamburg,
Physics Department,
Institut f\"ur Experimentalphysik,
Luruper Chaussee 149,
D-22761 Hamburg, Germany
}

\author{\centering 
K.\,Briggl, 
P.\,Eckert, 
T.\,Harion, 
H.\,-Ch.\,Schultz-Coulon, 
W.\,Shen, 
R.\,Stamen
 \\ \it
 University of Heidelberg, Fakult\"at f\"ur Physik und Astronomie,
Albert \"Uberle Str. 3-5, 2.OG Ost,
D-69120 Heidelberg, Germany
}

\author{\centering 
B.\,Bilki$^j$, 
E.\,Norbeck$^b$, 
D.\,Northacker,
Y.\,Onel
 \\ \it
University of Iowa, Dept.\ of Physics and Astronomy,
203 Van Allen Hall, Iowa City, IA 52242-1479, USA
}

\author{\centering 
G.\,W.\,Wilson
 \\ \it
University of Kansas, Department of Physics and Astronomy,
Malott Hall, 1251 Wescoe Hall Drive, Lawrence, KS 66045-7582, USA
}

\author{\centering 
K.\,Kawagoe,
Y.\,Sudo,
T.\,Yoshioka
 \\ \it
Department of Physics, Kyushu University, Fukuoka 812-8581, Japan
}

\author{\centering 
P.\,D.\,Dauncey
 \\ \it
Imperial College London, Blackett Laboratory,
Department of Physics,
Prince Consort Road,
London SW7 2AZ, UK 
}

\author{\centering 
M.\,Wing
 \\ \it
Department of Physics and Astronomy, University College London,
Gower Street,
London WC1E 6BT, UK
}

\author{\centering 
F.\,Salvatore$^k$ 
 \\ \it
Royal Holloway University of London,
Department of Physics,
Egham, Surrey TW20 0EX, UK
}

\author{\centering 
E.\,Cortina Gil, 
S.\,Mannai
 \\ \it
Center for Cosmology, Particle Physics and Cosmology (CP3)
Universit\'{e} catholique de Louvain, Chemin du cyclotron 2,
1320 Louvain-la-Neuve, Belgium
}

\author{\centering 
 G.\,Baulieu, 
 P.\,Calabria, 
 L.\,Caponetto, 
 C.\,Combaret, 
 R.\,Della\,Negra, 
 G.\,Grenier, 
 R.\,Han, 
 J-C.\,Ianigro, 
 R.\,Kieffer, 
 I.\,Laktineh, 
 N.\,Lumb, 
 H.\,Mathez, 
 L.\,Mirabito, 
 A.\,Petrukhin, 
 A.\,Steen, 
 W.\,Tromeur, 
 M.\,Vander\,Donckt, 
 Y.\,Zoccarato 
  \\ \it
Universit\'{e} de Lyon, Universit\'{e} Lyon 1, 
CNRS/IN2P3, IPNL 4, rue E.\ Fermi, 69622
Villeurbanne CEDEX, France
}

\author{\centering 
E.\,Calvo~Alamillo, 
M.-C.\, Fouz, 
J.\,Puerta-Pelayo 
 \\ \it
CIEMAT, Centro de Investigaciones Energeticas, Medioambientales y Tecnologicas, Madrid, Spain 
}

\author{\centering 
F.\,Corriveau
 \\ \it
Institute of Particle Physics of Canada and Department of Physics
Montr\'{e}al, Quebec,
Canada H3A 2T8
}

\author{\centering 
B.\,Bobchenko, 
M.\,Chadeeva, 
M.\,Danilov$^l$, 
A.\,Epifantsev, 
O.\,Markin, 
R.\,Mizuk$^l$, 
E.\,Novikov, 
V.\,Popov, 
V.\,Rusinov, 
E.\,Tarkovsky
 \\ \it
Institute of Theoretical and Experimental Physics, B. Cheremushkinskaya ul. 25,
RU-117218 Moscow, Russia
}

\author{\centering 
N.\,Kirikova, 
V.\,Kozlov, 
P.\,Smirnov, 
Y.\,Soloviev
 \\ \it
P.\,N.\, Lebedev Physical Institute,
Russian Academy of Sciences,
117924 GSP-1 Moscow, B-333, Russia
}

\author{\centering 
D.\,Besson, 
P.\,Buzhan,
A.\,Ilyin, 
V.\,Kantserov, 
V.\,Kaplin, 
A.\,Karakash, 
E.\,Popova, 
V.\,Tikhomirov
\\\it
Moscow Physical Engineering Inst., MEPhI,
Dept.\ of Physics,
31, Kashirskoye shosse,
115409 Moscow, Russia
}

\author{\centering 
C.\,Kiesling, 
K.\,Seidel, 
F.\,Simon, 
C.\,Soldner, 
M.\,Szalay, 
M.\,Tesar, 
L.\,Weuste 
 \\ \it
Max Planck Inst.\ f\"ur Physik,
F\"ohringer Ring 6,
D-80805 Munich, Germany
}

\author{\centering 
M.\,S.\,Amjad, 
J.\,Bonis, 
S.\,Callier, 
S.\,Conforti\,di\,Lorenzo, 
P.\,Cornebise, 
Ph.\,Doublet, 
F.\,Dulucq, 
J.\,Fleury, 
T.\,Frisson, 
N.\,van der Kolk,  
H.\,Li$^m$, 
G.\,Martin-Chassard, 
F.\,Richard, 
Ch.\,de la Taille, 
R.\,P\"oschl, 
L.\,Raux, 
J.\,Rou\"en\'e, 
N.\,Seguin-Moreau
 \\ \it
Laboratoire de l'Acc\'{e}l\'{e}rateur Lin\'{e}aire, Centre
Scientifique d'Orsay, Universit\'{e} de Paris-Sud XI, CNRS/IN2P3, BP
34, B\^atiment 200, F-91898 Orsay CEDEX, France
}

\author{\centering 
 M.\,Anduze, 
 V.\,Balagura, 
 V.\,Boudry, 
 J-C.\,Brient, 
 R.\,Cornat, 
 M.\,Frotin, 
 F.\,Gastaldi,  
 E.\,Guliyev$^n$, 
 Y.\,Haddad, 
 F.\,Magniette, 
 G.\,Musat, 
 M.\,Ruan$^o$, 
 T.H.\,Tran, 
 H.\,Videau
 \\ \it
 Laboratoire Leprince-Ringuet (LLR)  -- \'{E}cole Polytechnique, CNRS/IN2P3, F-91128 Palaiseau, France
}

\author{\centering 
B.\,Bulanek, 
J.\,Zacek 
 \\ \it
Charles University, Institute of Particle \& Nuclear Physics,
V Holesovickach 2,
CZ-18000 Prague 8, Czech Republic  
}

\author{\centering 
J.\,Cvach, 
P.\,Gallus, 
M.\,Havranek, 
M.\,Janata, 
J.\,Kvasnicka, 
D.\,Lednicky, 
M.\,Marcisovsky,  
I.\,Polak, 
J.\,Popule, 
L.\,Tomasek, 
M.\,Tomasek, 
P.\,Ruzicka, 
P.\,Sicho, 
J.\,Smolik, 
V.\,Vrba, 
J.\,Zalesak 
 \\ \it
Institute of Physics, Academy of Sciences of the Czech Republic, Na Slovance 2,
CZ-18221 Prague 8, Czech Republic
}

\author{\centering 
B.\,Belhorma, 
H.\,Ghazlane 
 \\ \it
Centre National de l'Energie, des Sciences et des Techniques Nucl\'{e}aires, 
B.P. 1382, R.P. 10001, Rabat, Morocco
}

\author{\centering              
K.\,Kotera, 
T.\,Takeshita, 
S.\,Uozumi
 \\ \it
Shinshu Univ.\,,
Dept. of Physics,
3-1-1 Asaki,
Matsumoto-shi, Nagano 390-861,
Japan
}

\author{\centering 
S.\,Chang, A.\,Khan, D.\,H.\,Kim, D.\,J.\,Kong, Y.\,D.\,Oh
\\ \it
Department of Physics, Kyungpook National University, Daegu, 702-701,
Republic of Korea
}

\author{{\centering 
M.\, G\"otze, 
J.\,Sauer, 
S.\,Weber, 
C.\,Zeitnitz
 \\ \it
Bergische Universit\"{a}t Wuppertal,
Fachbereich C Physik,
Gaussstrasse 20,
D-42097 Wuppertal, Germany
}
 \\ \it
$^\spadesuit$ Corresponding author\newline
E-mail: \email{angela.isabela.lucaci.timoce@cern.ch}
}

\author{  \\
\llap{$^a$}Also at University of Texas, Arlington\\
\llap{$^b$}Deceased\\
\llap{$^c$}Now at DESY Hamburg, Germany\\
\llap{$^d$}Now at University of Bergen, Norway\\
\llap{$^e$}Now at Max Planck Inst. f\"ur Physik, Munich, Germany\\
\llap{$^f$}Now at Stony Brook University (SUNY), Dept.\ of Physics and
Astronomy, Stony Brook, NY, USA\\
\llap{$^g$}Now at Hamburg University, Germany \\
\llap{$^h$}On leave from ITEP\\
\llap{$^i$}Also at DESY Hamburg, Germany\\
\llap{$^j$}Also at Argonne National Laboratory\\
\llap{$^k$}Now at University of Sussex, Physics and Astronomy
Department, Brighton, Sussex, BN1 9QH, UK\\
\llap{$^l$}Also at MEPhI and at Moscow Institute of Physics and Technology\\
\llap{$^m$}Now at LPSC Grenoble\\
\llap{$^n$}TRIUMF, Vancouver, BC, Canada\\
\llap{$^o$}Now at IHEP, Beijing, China

}

\abstract{Lepton colliders are considered as options to complement and
to extend the physics programme at the Large Hadron Collider. The
Compact Linear Collider (CLIC) is an $e^+e^-$ collider under
development aiming at
centre-of-mass energies of up to 3~TeV. For experiments at CLIC, a hadron
sampling calorimeter with tungsten absorber is proposed. Such a
calorimeter provides sufficient depth to contain high-energy showers,
while allowing a compact size for the surrounding solenoid.

A fine-grained calorimeter prototype with tungsten absorber plates and
scintillator tiles read out by silicon photomultipliers was built and
exposed to particle beams at CERN. Results obtained with electrons,
pions and protons of momenta up to 10~GeV are presented in terms of
energy resolution and shower shape studies. The results are compared
with several GEANT4 simulation models in order to assess the
reliability of the Monte Carlo predictions relevant for a future
experiment at CLIC.}


\begin{document}
\section{Introduction}
The Compact Linear Collider (CLIC) is a possible future $e^+e^-$
collider~\cite{CLIC_CDR} that would allow the exploration of a new energy region in
the multi-TeV range, beyond the capabilities of today's particle
accelerators.
The main driver for the design of the CLIC detector concept is the
requirement for a jet energy resolution close to $30\%/\sqrt{E\:[\mathrm{GeV}]}$. This can be
achieved by using fine-grained calorimeters and particle-flow analysis
techniques~\cite{Thomson}.
Simulation studies showed that a dense material has to be used as
absorber in the calorimeter, in order to contain the high-energy
showers, while limiting the diameter of the surrounding
solenoid. The detector concepts being developed for CLIC feature a
barrel calorimeter with tungsten absorber plates.

In order to test such a detector, the CALICE
collaboration~\cite{CALICE_web} constructed a tungsten absorber structure, to be
combined with existing readout layers of the Analog Hadron Calorimeter
(\mbox{AHCAL})~\cite{CALICE_AHCAL}. Data were recorded with the CALICE tungsten AHCAL
(W-AHCAL) prototype  at the CERN PS in September-October 2010 with
mixed beams containing muons, electrons, pions and
protons in the momentum range of 1 to 10~GeV/c.
This paper presents energy resolution measurements  and
studies of the longitudinal and radial shower development. 

In section~\ref{sec:experimentalSetup} we briefly describe the
experimental setup. The procedure to calibrate the calorimeter and the
temperature corrections are presented in section~\ref{sec:calibration}.
  Section~\ref{sec:simulation} introduces details about the Monte
  Carlo simulation. The systematic uncertainties are discussed in section~\ref{sec:sys}.
In sections~\ref{sec:electronAnalysis}, \ref{sec:hadronAnalysis} and
\ref{sec:response} the analyses of the electron, pion and
proton data and comparisons to the Monte Carlo simulations are
presented. A summary of the results is given in
section~\ref{sec:summary}.

\section{Experimental setup}
\label{sec:experimentalSetup}

\begin{figure}[t!]
\centering
\includegraphics[width=0.2\textwidth, angle=-90]{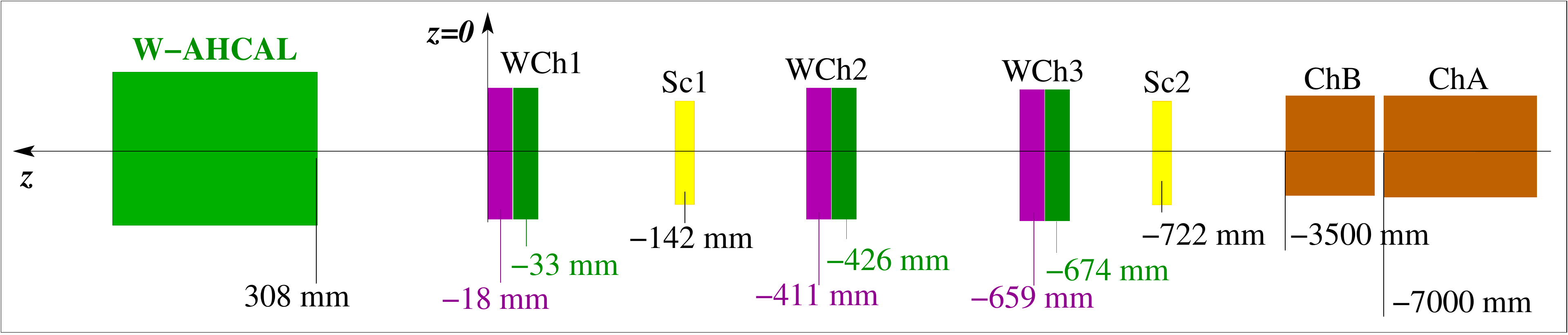}
\caption{Sketch of the experimental setup at the T9
beam line of the CERN PS (not to scale).
   $Sc$ stands for $scintillator$, $WCh$ for $wire\; chamber$, and
   $Ch$ for $Cherenkov$. The beam enters from the right.}
\label{fig:CERN_PS_2010_testBeamLine}
\end{figure}

The W-AHCAL consists of a 30-layer sandwich structure of
absorber plates interleaved with 0.5~cm thick scintillator tiles,
read out by wavelength shifting fibres coupled to silicon
photo-multipliers (SiPMs). The calorimeter has a total
of 6480 channels. One absorber plate is 1~cm thick and is made of a
tungsten alloy consisting of 92.99\% tungsten, 5.25\% nickel and
1.76\% copper, with a density of
17.8~g/cm$^3$. The nuclear interaction length
of this alloy
is $\lambda_\mathrm{I} = 10.80$~cm and the radiation length is
$X_0=0.39$~cm. 
The scintillator tiles are placed into a steel
cassette, with 0.2~cm thick walls. Thus one calorimeter layer
corresponds to 0.13~$\lambda_{\mathrm{I}}$ and to 2.8~$X_0$.
 The overall dimensions of the prototype
are $0.9\times 0.9\times 0.75\;\mathrm{m}^3$,
amounting to 3.9~$\lambda_{\mathrm{I}}$ and to
85~$X_0$. 
The high granularity of the detector is
ensured by  the $3\times 3\; \mathrm{cm}^2$ tiles placed in the centre of each active plane, 
 surrounded by $6\times 6\;
\mathrm{cm}^2$ and \mbox{$12\times 12\;\mathrm{cm}^2$} tiles at the
edges.  Since the SiPM response varies with
the temperature, the latter is monitored
in each layer by 5~sensors~\cite{CALICE_AHCAL}.

The data were recorded in the secondary T9
beam line~\cite{SPS-T9-beamline} of the
CERN PS East Area~\cite{CERN-PS-east-area}.
The 24~GeV/c primary proton beam hits a  target 57~m upstream of the W-AHCAL prototype.
A momentum-selection and focusing system is used to deliver a
mixed beam of electrons, muons, pions and protons with momenta
between 1 and 10~GeV/c.
The momentum spread 
$\Delta p$/$p$ is  of the order of 1\% for all momenta. The beam size is chosen such that
the resulting Gaussian
spread
on the \mbox{W-AHCAL} surface is approximately $3\times3$~cm$^2$ for 10~GeV/c pions.

A sketch of the CERN PS test beam setup is presented in
figure~\ref{fig:CERN_PS_2010_testBeamLine}. The secondary beam passes two Cherenkov
threshold counters (A and B), two trigger scintillators and a tracking
system of three wire chambers.
The Cherenkov counters are filled with $\mathrm{CO}_2$ gas
 with pressures adjustable up to 3.5~bar.
The Cherenkov information is read out through photo-multiplier tubes
and subsequent discriminators with a fixed threshold.
The Cherenkov signals are used offline for particle identification.
The beam trigger is defined by the coincidence of two
$10\times10\times1$~cm$^3$ scintillator counters. The information from three $11\times11$~cm$^2$ wire
chambers~\cite{wire_chambers} is used offline to reconstruct the track of the incident particle
and predict its position on the calorimeter surface.

\begin{figure}[t!]
\centering
\includegraphics[scale=0.3]{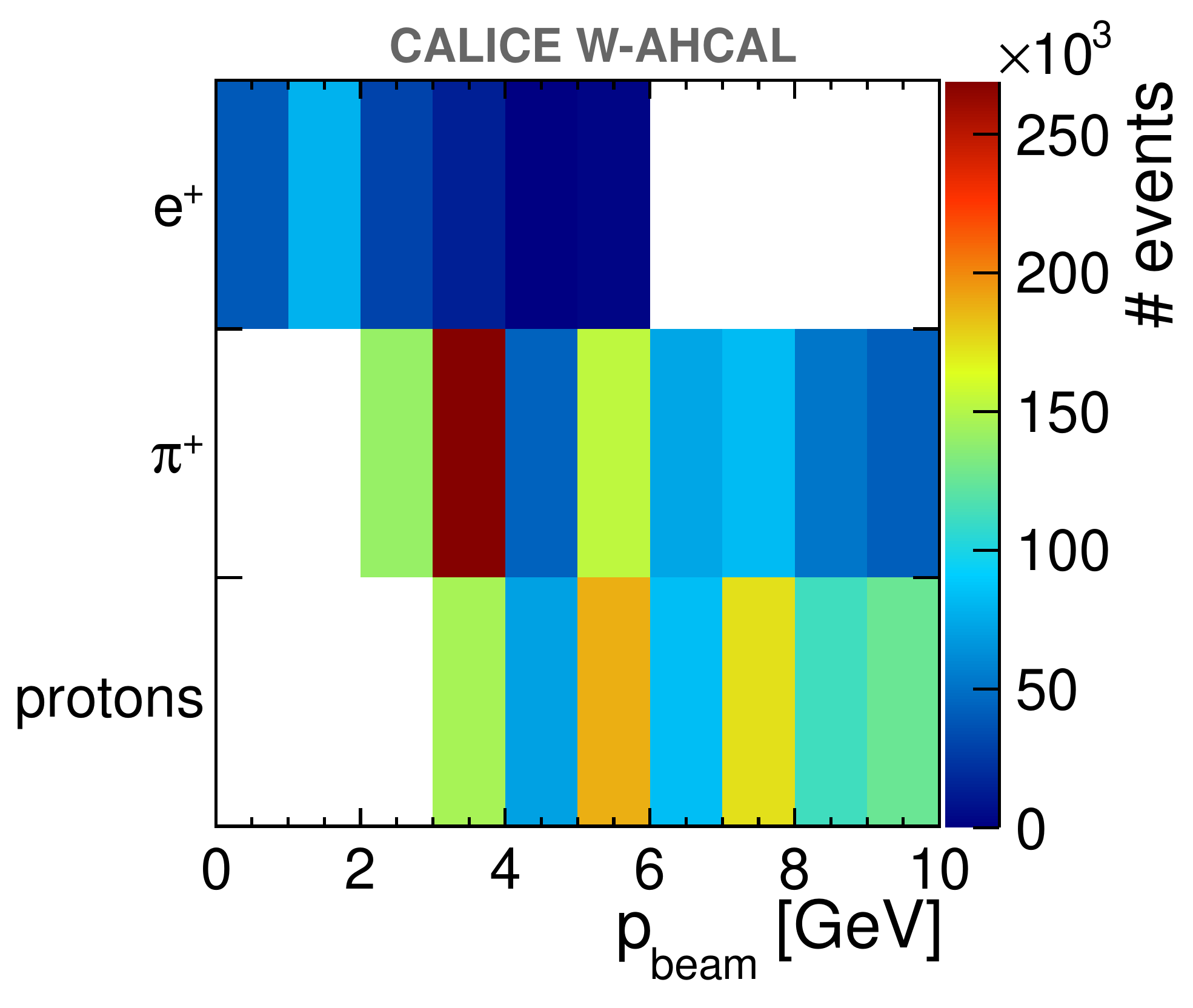}
\caption{Number of events after the selection of a given particle type
from the positive-polarity data.}
\label{fig:nevents}
\end{figure}

The data recorded by the CALICE W-AHCAL in 2010
contained a mixed beam of particles.
 The negative-polarity beam contains
$e^-$, $\pi^-$ and $\mu^-$ particles. The anti-proton content was
considered to be negligible. The positive-polarity beam contains
$e^+$, $\pi^+$, $\mu^+$ and protons. The kaon content was negligible
for both polarities.
The distribution of the number of events after the selection of a
given particle type  from the positive-polarity data is given in
figure~\ref{fig:nevents}. The numbers for the negative-polarity beams
are similar.

\section{Calibration and temperature correction}
\label{sec:calibration}

The responses of all calorimeter cells are calibrated to a common
physics signal based on minimum ionising particles (MIP) which
were obtained in dedicated muon runs. Several steps are
necessary to translate the signals measured with the SiPM readout (in ADC
counts) to information about the deposited energy (in MIP).

The calibration of
a single cell $i$ is done according to the formula:
\begin{equation}
\displaystyle E_i [\mathrm{MIP}] =
\frac{A_i[\mathrm{ADC}]}{A_i^{\mathrm{MIP}}[\mathrm{ADC}]}\cdot
f_{\mathrm{resp}}(A_i[\mathrm{pixels}]),
\label{eq:calibFormula}
\end{equation}
where:
\begin{itemize}
\item $A_i[\mathrm{ADC}]$ is the pedestal-subtracted amplitude registered in cell $i$, in
  units of ADC counts;
\item $A_i^{\mathrm{MIP}}[\mathrm{ADC}]$ is the pedestal-subtracted MIP amplitude in cell
  $i$, measured in ADC counts. It is taken as the most probable
  value of the energy response for muons;
\item $f_{\mathrm{resp}}(A_i[\mathrm{pixels}])$ is
  the SiPM saturation correction
  function which corrects for the
  non-linearity of the SiPM response. 
  This function assumes an
  effective number of total pixels of about 925. 
\end{itemize}

  The amplitude in units of pixels is obtained by dividing the amplitude of a
  cell by the corresponding SiPM gain factor $G_i[\mathrm{ADC}]$:
  \begin{equation}
    \label{eq:gain}
    \displaystyle A_i[\mathrm{pixels}] = \frac{A_i[\mathrm{ADC}]}{G_i[\mathrm{ADC}]}.
  \end{equation}
   The gain values are obtained from fits of photo-electron spectra
   taken with low intensity LED light provided by a calibration and
   monitoring LED system. 
Detailed information about the calibration and the saturation
  correction procedures can be found
in~\cite{CALICE_AHCAL_emPaper}. After calibration, only cells with an
  energy above 0.5~MIP are considered, in order to reduce the noise contribution.

During data taking, the SiPM noise
spectra  were monitored to identify channels which give
no signal, or which give too high a signal. 
These types of
channels are identified based on the RMS value of the energy
distributions from dedicated random trigger runs:
\begin{itemize}
\item Dead channels: RMS < 20.5 ADC counts.
\item Noisy channels: RMS > 140 ADC counts.
\end{itemize}
On average, during the CERN 2010 data taking period less than 3\% of the
total number of calorimeter channels were identified as noisy or dead,
and discarded from the analysis.

 As the SiPM response depends on temperature, only muon runs within a narrow temperature
range ($T=25.0 \pm 0.5^\circ$~C) were used for measuring the
$A_i^{\mathrm{MIP}}[\mathrm{ADC}]$ calibration
constants. From the total of 6480~channels, 92\% had sufficient
statistics and the corresponding $A_i^{\mathrm{MIP}}[\mathrm{ADC}]$
calibration factors were determined.
The other channels were discarded from the analysis.

The temperature inside the calorimeter is measured by 5~sensors for each calorimeter layer. 
The sensors are horizontally centred within the layer and equally
spaced vertically. The
vertical temperature spread was found to be of the order of
$0.5^{\circ}\:\mathrm{C}$ per plane. The average calorimeter temperature for the analysed runs varied from
20 to 25$^\circ$~C.

\begin{figure}[t!]
\begin{minipage}[t]{0.48\linewidth}
\centering
\includegraphics[scale=0.35]{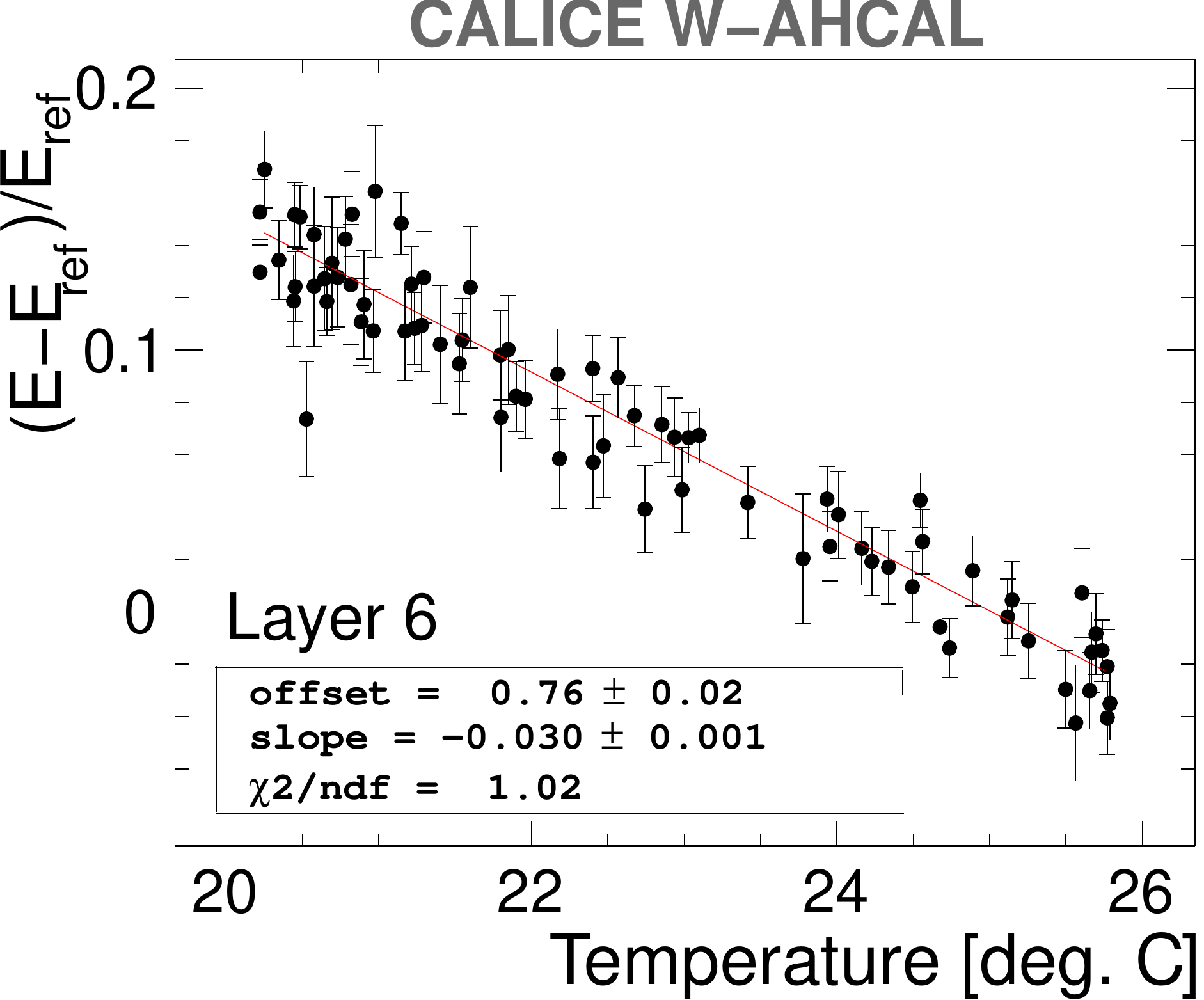}
\caption{Example of the measurement of the MIP temperature dependence for
the \mbox{W-AHCAL} layer~6. Each data point corresponds to the most
 probable value of the energy response
 in a given run of all calorimeter cells in this layer that belonged to
 a muon track.
}
\label{fig:mipExampleRelSlope}
\end{minipage}
\hspace{0.5cm}
\begin{minipage}[t]{0.48\linewidth}
\centering
\includegraphics[scale=0.35]{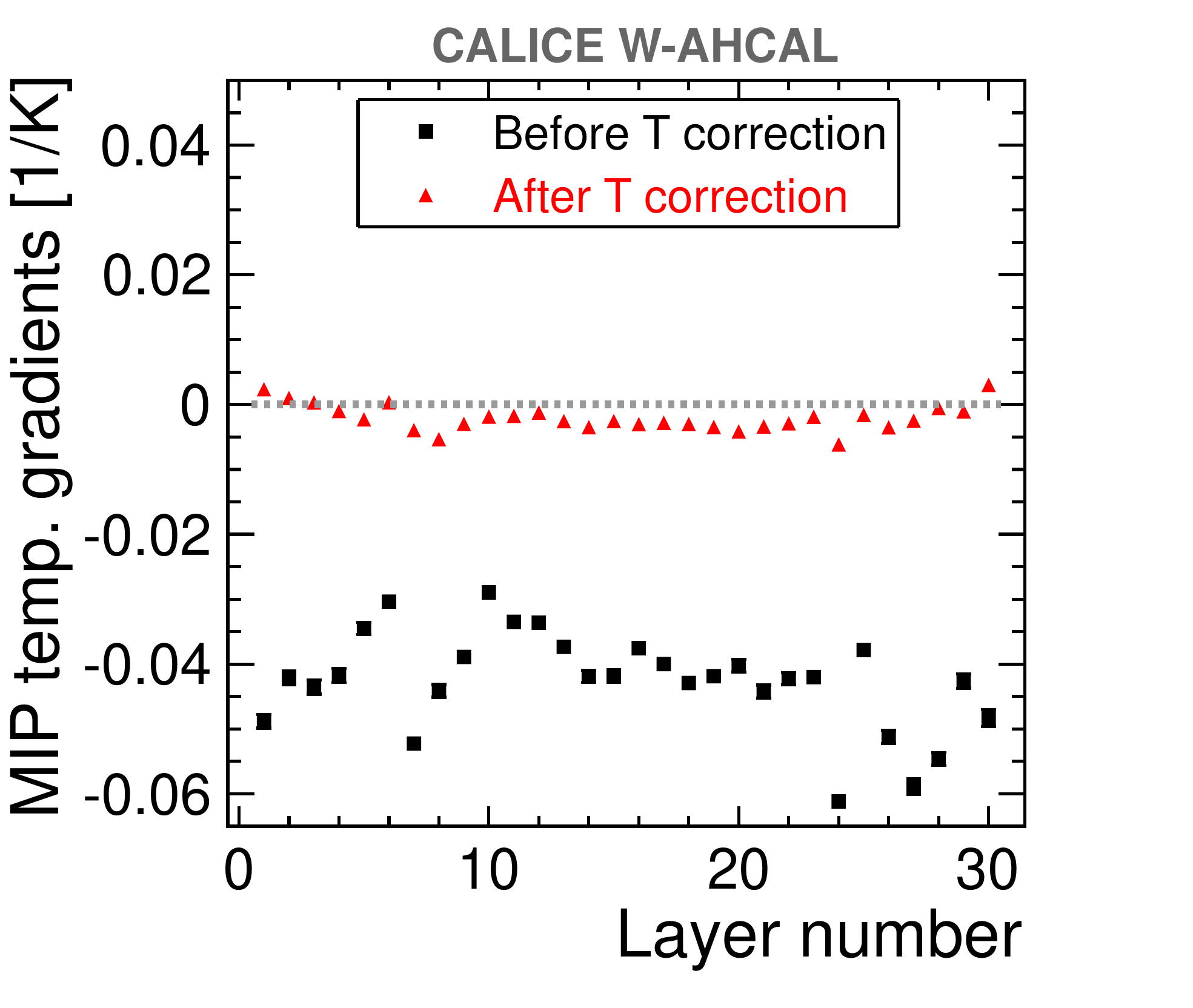}
\caption{Distribution of the linear MIP temperature
 gradients
per \mbox{W-AHCAL} layer,
  before and after temperature correction. The average gradient
  is $-4.3\%$/K before the correction, and \mbox{$-0.2\%$/K} after.}
\label{fig:mipRelSlopes}
\end{minipage}
\end{figure}

The MIP calibration factors show an inverse linear dependence on
temperature. 
Therefore, in order to take into account the possible temperature differences between
the muon calibration runs and the analysed data runs, the MIP
calibration factors are corrected for the temperature differences.
To measure the temperature dependence, muon tracks are identified using a track
finder. Then the position of the most probable value was found using the energy distribution of all muon
track hits in a given layer. The dependence of the peak position
on the temperature is fitted with a linear function for
each calorimeter layer, as
illustrated in figure~\ref{fig:mipExampleRelSlope}. 
The linear dependence, expressed in percent
per Kelvin, is measured relative to the calorimeter response $E_{\mathrm{ref}}$
obtained at the temperature ($T=25.0 \pm 0.5^\circ$~C) quoted above,
at which the muon calibration runs were
taken. The distributions of the MIP temperature slopes per W-AHCAL
layer, before and after temperature correction, are shown in
figure~\ref{fig:mipRelSlopes}. After correction, the average slope is
at the level of $-0.2\%$/K.  The remaining
temperature gradients are due to uncertainties in the
temperature measurements within a calorimeter layer.

\section{Monte Carlo simulation}
\label{sec:simulation}

A simulation of the experimental setup is implemented in a GEANT4~\cite{GEANT4} based
application~\cite{Mokka}. The simulated geometry includes the
full setup starting from 60~m
upstream of the calorimeter with the  scintillators, the wire
chambers and the W-AHCAL.
The beam position and spread are measured using the information from
the wire chambers and included in the simulation of each run.

\subsection{GEANT4 models}
\label{sec:GEANT4models}

 The physics models in the GEANT4 simulation are combined into so-called
physics lists, providing a balance between the level of physics
precision and CPU performance. Within a list, the models are valid in
different energy ranges and for different particles.

\begin{figure}[t!]
\centering
\includegraphics[width=0.6\textwidth]{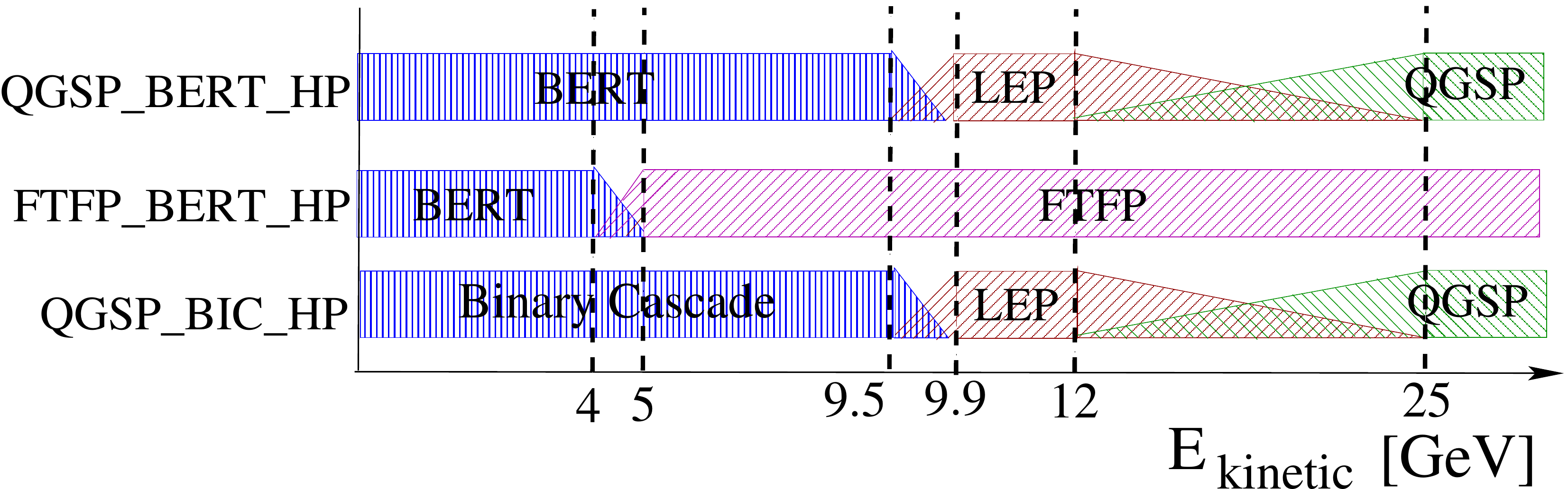}
\caption{Schematic representation of selected GEANT4 physics lists
with the energy ranges of the different models. In the overlap regions between the models, a
  random choice between the corresponding models is performed, based
  on the kinetic energy of the incident particle in each interaction.}
\label{fig:physLists}
\end{figure}

Several GEANT4 (version 9.6.p02) physics lists were selected in order
to compare them with the hadron data: 
\begin{itemize}
\item \textbf{QGSP\_BERT\_HP}: Employs the Bertini (BERT) cascade
model which handles incident nucleons, pions and kaons with kinetic
energy up to 9.9~GeV. From 9.5 to 25~GeV it uses the low energy
parametrised (LEP) model. For energies above 12~GeV it employs the
quark-gluon string precompound (QGSP) model.

\item \textbf{QGSP\_BIC\_HP}: Uses the binary cascade (BIC) model for  
incident protons and neutrons with a kinetic energy
  \mbox{$E_{\mathrm{kin}} < 10$~GeV} and  pions with
  $E_{\mathrm{kin}} < 1.5$~GeV. For energies above 9.9~GeV this physics list is
  identical to QGSP\_BERT\_HP.

\item \textbf{FTFP\_BERT\_HP}: Uses the Bertini cascade model up to
  5~GeV, then the Fritiof precompound (FTFP) model.
\end{itemize}
A more detailed description can be found for
example in~\cite{GEANT4_physLists}.
 The
combinations of the simulations models for selected physics lists
are presented in figure~\ref{fig:physLists}.

As the W-AHCAL uses tungsten as absorber material, slow neutrons are expected to play
an important role in hadron interactions in this calorimeter. 
Therefore the above-mentioned physics lists are combined with the
data-driven Neutron High Precision (HP) Models and Cross Sections, which
treat the detailed simulation of the interaction, transportation,
elastic scattering and capture of neutrons with energies below 20~MeV. 
Since the electromagnetic model is the same for all
GEANT4 physics lists, the $e^{\pm}$ data are compared with the
$\mathrm{QGSP\_BERT\_HP}$ physics list only.

\subsection{Generation and digitisation of the simulation}
\label{sect:digitization}

Events are generated for each of the selected physics
lists described in section~\ref{sec:GEANT4models}. 
To compare simulation with data, one needs to consider realistic
detector effects which occur in addition to the particle
interaction and energy deposition. This is done both at the generation
and digitisation level.

At the generation step, the following aspects are taken into
account:
\begin{itemize}
\item \textbf{Signal shaping time of the readout electronics}: To
  emulate the signal shaping time, only hits within a time window of 150~ns (corrected
  for the time of flight) are accepted. The start of the time window is defined
  from the moment when the particle reaches the W-AHCAL front face.
\item \textbf{Non-linearity of the light output}: In the case of
  plastic scintillator, the light output per unit length has a
  non-linear dependence on the energy loss per unit length of the
  particle's track. This behaviour is described by the so-called
  \textit{Birks' law}~\cite{Birks}:
  \begin{equation}
    \displaystyle \frac{dL}{dx} \propto \frac{dE/dx}{1+k_{\mathrm{Birks}}\cdot dE/dx},
  \end{equation}
  where $dL$/$dx$ represents the light output per unit length, $dE$/$dx$
  is the energy lost by the particle per unit length of its path (in
  units of MeV/mm), and
  $k_{\mathrm{Birks}}$ is a material-dependent factor (Birks constant). The Birks' law
  is applied to the W-AHCAL hits, using a factor of $k_{\mathrm{Birks}}=0.07943$~mm/MeV~\cite{Hirschberg}.
\end{itemize}

For the digitisation of the signal, the
same sets of calibration values and of dead or uncalibrated channels
are used  as for the reconstruction of the experimental data.
In a first digitisation step, the simulated energy (in GeV) is
converted into MIP based on a MIP-to-GeV factor obtained with
simulated muons.
Next, the following aspects are taken into account: the detector granularity, the
light sharing between the tiles, the non-linear SiPM response due to
saturation and the conversion of the signal from MIP to ADC counts, 
the statistical smearing of the detector response at the pixel
scale, and the contribution from electronic noise (obtained from
data). 
At this stage, the energy of the simulated hits is expressed in ADC
counts, and is given as input to the same calibration procedure as 
 for
the data (see equation~\ref{eq:calibFormula}).

\section{Systematic uncertainties}
\label{sec:sys}

\subsection{Systematic uncertainties for data}
The main contributions to the systematic uncertainties that affect the
measurement of the total calorimeter response were found to be:
\begin{itemize}
\item $\pm 2\%$ uncertainty due to the MIP calibration factors. These
factors are obtained by fitting the muon hit energy distributions for
a given cell. The fit results depend on the fitting function, on the
binning of the histograms, and on the muon track selection, resulting
into an overall uncertainty of $\pm 2\%$ on the MIP calibration factors.

\item $\pm1.5\%$ due to the run-wise variation of the calorimeter
response. This value is given by the RMS of the
mean energy sum in the analysed runs (without subtracting the
contribution due to the statistical error on each mean).
\end{itemize}

In addition, the electromagnetic showers are affected by the
uncertainties in the measurement of the saturation scaling factor.   
This was studied by randomly varying the saturation
scaling factors according to a Gaussian distribution with a mean of 0.8, and
a sigma of 0.09, as obtained in dedicated
measurements~\cite{CALICE_AHCAL_emPaper}.
The  data were calibrated successively 100~times
using the smeared scaling factors. With this method, variations of the average
total energy deposited in the calorimeter ranging from 0.1\% for
1~GeV positrons, to 0.4\% for 6~GeV positrons, were obtained. This
effect is greatly reduced for hadron-induced showers which typically
have a larger number of active cells.

The uncertainty on the measurement of the energy per calorimeter layer
due to the channel-wise variation of the MIP calibration factors 
affects the longitudinal profiles, because a signal in an individual layer can be
dominated by a few cells, as in the case of electromagnetic
showers. This uncertainty was determined from the width of the
distribution of the difference between the MIP calibration factors
measured in two independent running periods, resulting in a variation of $\pm 3.6\%$.

\subsection{Systematic uncertainties for the Monte Carlo simulation}

Due to the imperfect reflective coating of the scintillator tiles,
light might leak between neighbouring calorimeter cells. This is taken
into consideration in the simulation via the so-called cross-talk
factor, which is the fraction of energy leaking into neighbouring
cells. Measurements of the cross-talk yielded values of
2.5\%~\cite{CALICE_AHCAL_emPaper} per tile edge.
Recent measurements in a different sample resulted in estimates between 3.3\% and
4.6\%. 
 To
account for the imperfect knowledge of the cross-talk, and hence of
the energy scale, an uncertainty of $+3\%$ is assumed conservatively for
the total energy sum in the simulation.

The impact of the integration time window of 150~ns, which is due to
the signal shaping time of the readout electronics, on the simulated
calorimeter response was also studied. Variation of $\pm30$~ns around
the time cut of 150~ns resulted in a negligible impact on the measured energies.

\section{Analysis of electron data}
\label{sec:electronAnalysis}

As the underlying physics of electromagnetic showers is
well understood, the analysis of the $e^\pm$ data is used to validate
the implementation of the detector material and response in the simulation, as well as
the calibration chain.
The electromagnetic analysis is also important for the study of the
degree of (non)compensation of the hadron calorimeter, which is
expressed in the $e/\pi$ ratio, i.e.\ the ratio 
of the detector response for electrons to that for hadrons.

\subsection{Data selection}
\label{sect:emSelection}

Only $e^{\pm}$ runs up to 6~GeV were considered;
  for higher energies, the $e^{\pm}$ content in the beam was too low.
The first level of selection is based on Cherenkov threshold counters~\cite{LCD-Open-2013-001}. 
Additional cuts were applied in order to reject the small
fraction (of the order of a few percent) of hadron and muon events in
the data sample.

While hadrons are expected to penetrate deep into the calorimeter, electrons start
to shower already in the first calorimeter layer, and most of the
shower is contained within the first five layers. To identify the
electromagnetic shower clusters
a nearest-neighbour algorithm~\cite{Lutz} is used.
Further a cut is applied
on the cluster centre-of-gravity in the $z$-direction,
$z_{\mathrm{cog}}^{\mathrm{cluster}}$, defined as:
\begin{equation}
\displaystyle z^{\mathrm{cluster}}_{\mathrm{cog}}= \frac{\sum_i E_i \cdot z_i}{\sum_i
  E_i},
\label{eq:zcog}
\end{equation}
where $z_i$ is the $z$-position of the cluster hits, and $E_i$ is
their energy.
Only events are selected which contain only one cluster that has the
centre-of-gravity along the beam axis in the first part of the
calorimeter, i.e.\ with \mbox{$z_{\mathrm{cog}}^{\mathrm{cluster}}<400$~mm},
which corresponds to approximately 3~calorimeter layers\footnote{The variable
$z_{\mathrm{cog}}^{\mathrm{cluster}}$ is calculated in the coordinates of the laboratory
frame, with the centre attached to the back plane of the first wire
chamber, 308~mm away from the W-AHCAL front face, as indicated in figure~\ref{fig:CERN_PS_2010_testBeamLine}.}.

To reduce the influence of noise in the $e^\pm$ events, only
calorimeter cells within the first 20~calorimeter layers and within
the central $10\times 10$ tiles of
$3\times3\;\mathrm{cm}^2$ are considered. This is safe because in all the
runs the beam profile is centred on the calorimeter centre,
and the width of the beam profile is not more than 3~tiles.

\begin{figure}[t!]
\centering
\includegraphics[width=0.45\textwidth]{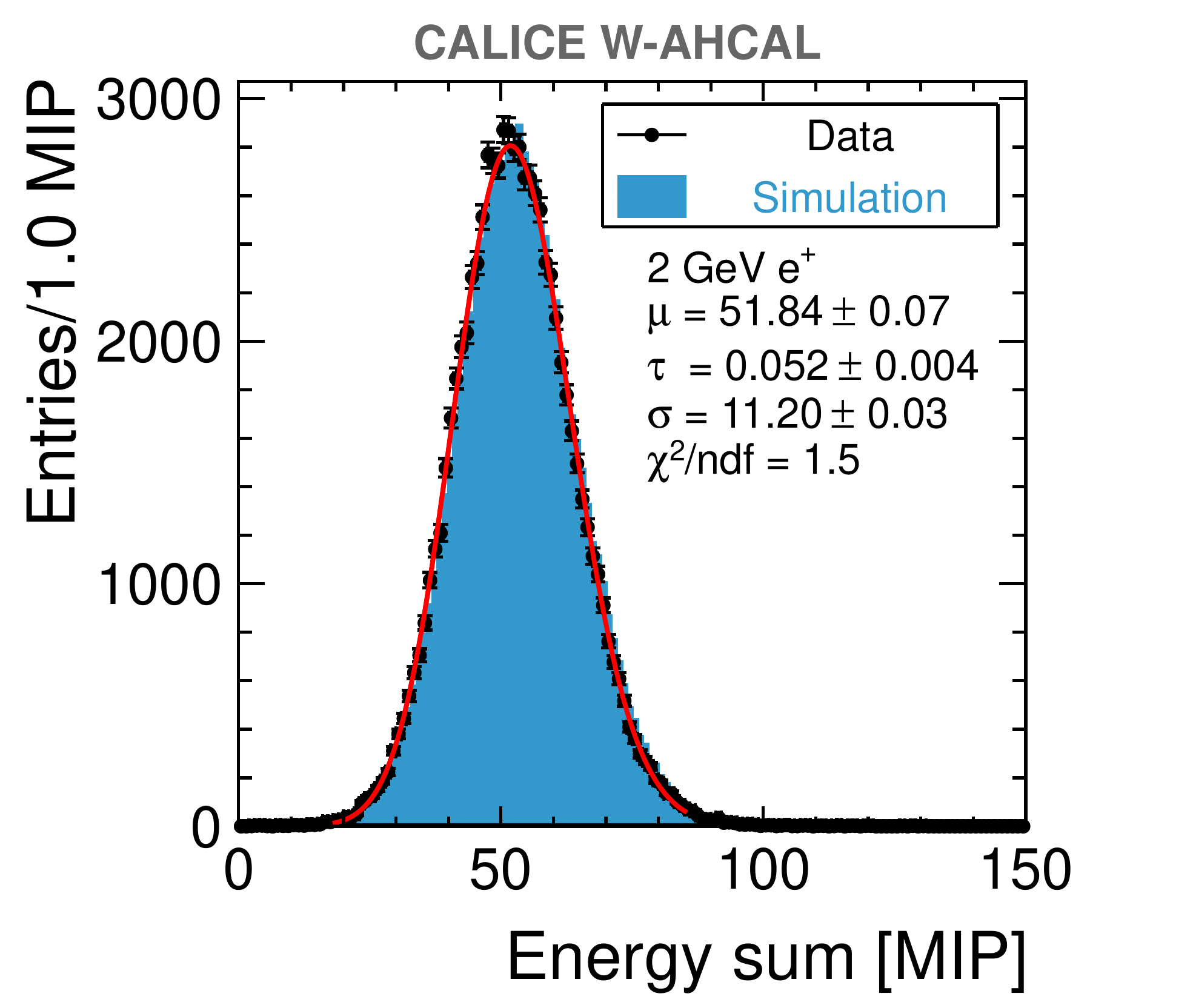}
\caption{The distribution of the energy sum deposited by 2~GeV~$e^+$
  in the CALICE W-AHCAL. Data, shown by the black filled circles, are
  compared to the simulation (filled area). The red line is the
  result of a fit of the data using the Novosibirsk function defined
  in equation~\protect\ref{eq:novo}. 
  The fit results are also
  given.}
\label{fig:ePlus_2GeV_fit_mc}
\end{figure}

The $e^{\pm}$ energy sum spectra have a non-Gaussian shape, with tails at
high energies, as can be seen for example for 2~GeV positrons
in figure~\ref{fig:ePlus_2GeV_fit_mc}. These high energy tails
 originate from the limited
number
 of active cells  in an electromagnetic shower, due to the 1~cm thick
 tungsten absorber per layer, which corresponds to 2.6 radiation lengths $X_0$.
On average, about
 17~cells are active in an electromagnetic shower induced by a 1~GeV
 particle, and about 38~cells in the 6~GeV case.
The energy spectra of individual cells, after
pedestal subtraction, are exponentially falling. With increasing number of
active cells, the total energy distribution becomes more and more
Gaussian\footnote{The central limit theorem states
  that the distribution of an average tends to be Gaussian for a large
  number of samples, even when
  the distribution from which the average is computed is decidedly
  non-Gaussian.}. The high energy tails are also present in the simulation, at
generator level, i.e.\ before including any detector effects.

The $e^\pm$ energy spectra are fitted with the Novosibirsk fit
function, which accounts for the high energy tails. 
 This function is
defined as~\cite{Novosibirsk}:
\begin{equation}
\label{eq:novo}
f(x) = A\cdot \exp \left[
-0.5\cdot \left(
\frac{\ln^2[1+\Lambda\cdot\tau\cdot(x-\mu)]}{\tau^2} + \tau^2
\right)
\right]
\end{equation}
where
\begin{equation}
\displaystyle \Lambda \equiv \frac{\sinh(\tau \cdot \sqrt{\ln
    4})}{\sigma \cdot \tau \cdot \sqrt{\ln 4}},
\end{equation}
with  $\mu$ the peak position, $\sigma$ the width, and $\tau$ the
tail parameter. 
With $\tau \rightarrow 0$ the function given in equation~\ref{eq:novo}
converges to a Gaussian with width~$\sigma$.
An example fit for 2~GeV positrons, together with the fit results, is
given in figure~\ref{fig:ePlus_2GeV_fit_mc}. The fit range is
$\pm3\sigma$ around the peak of an initial fit with the same
function.  The $\mu$ parameter gives the mean energy sum in the
W-AHCAL, further denoted by $\langle E_{\mathrm{vis}}\rangle$, i.e.\
visible energy. 
It was checked that the $\mu$ parameter from the
  Novosibirsk fit is the same as the statistical mean of the
  distribution within uncertainties.
The $\sigma$ parameter gives the width of the
distribution, and it is further used to measure the $e^+$ energy
resolution.

\subsection{Electromagnetic response and energy resolution}

\begin{figure}[t!]
\begin{minipage}[c]{0.48\textwidth}
\centering
\includegraphics[scale=0.31]{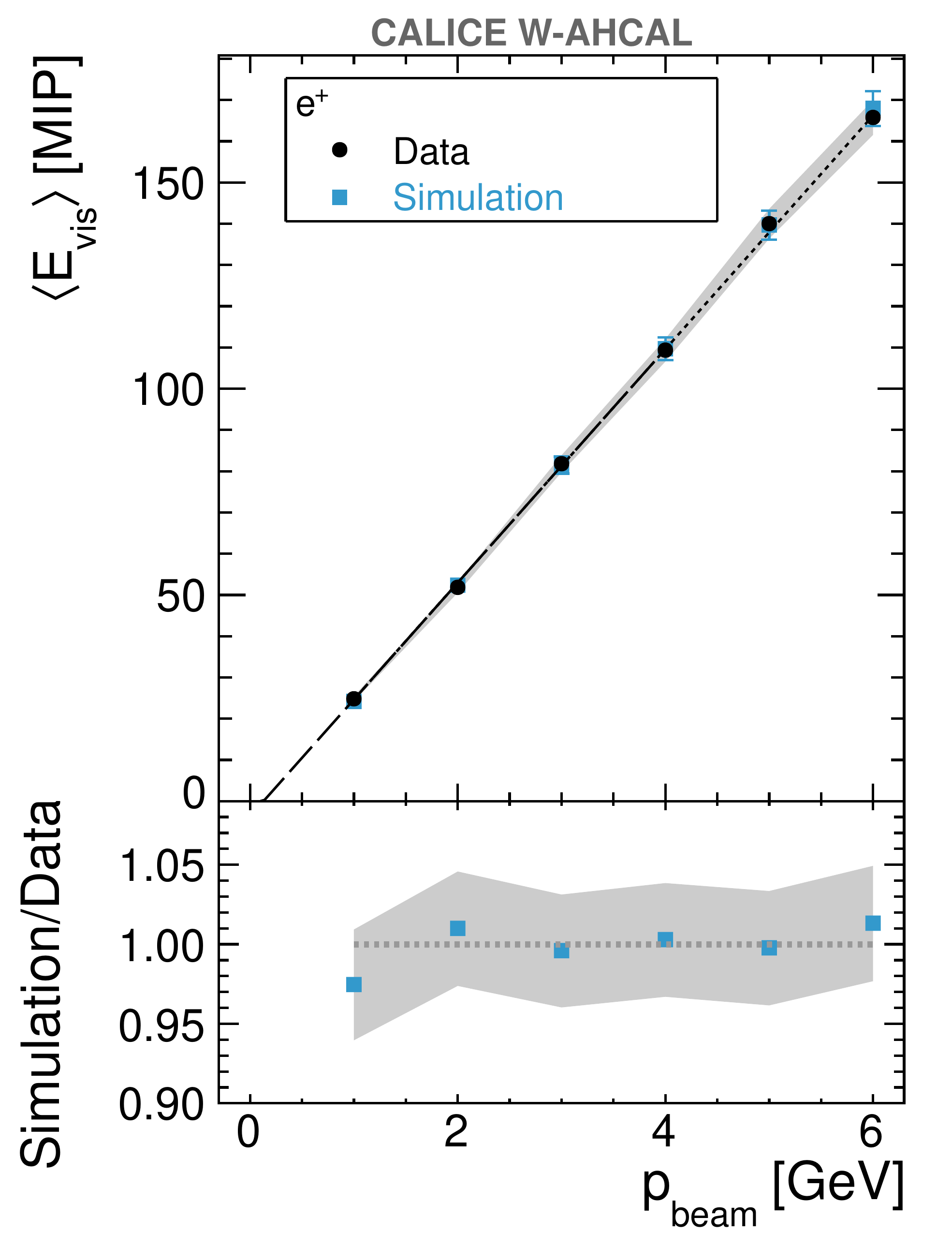}
\caption{Dependence of the mean visible positron energy on the  beam momenta.
  The data are compared  with the simulation.  The line indicates a
  fit with the
  function \mbox{$\langle E_{\mathrm{vis}}\rangle =u + v\cdot
  p_{\mathrm{beam}}$}. In the bottom part,
  the ratio between the simulation and the data is shown. The grey
  band shows the overall uncertainty for both data and simulation.}
\label{fig:ePlus_linearity_mc}
\end{minipage}
\hspace{0.5cm}
\begin{minipage}[c]{0.48\textwidth}
\centering
    \captionof{table}{Fit parameters of the  dependence of the mean
    positron visible energy on the  beam momentum: comparison of
    data with simulation.}
    \label{tab:ePlus_linearity_mc_results}
    \begin{tabular}{lll}\hline
      Parameter &  Data & Simulation \\\toprule
      $u$ [MIP] & $-3.67 \pm 0.92$ & $-4.56 \pm 2.01$\\
      $v$ [MIP/GeV] &  $28.32 \pm 0.49$ &  $28.68 \pm 1.07$\\
      $\chi^2$/ndf & 1.3/4 & 0.2/4\\\bottomrule
      \end{tabular}
\end{minipage}
\end{figure}

The calorimeter response for electromagnetic showers is expected
to be linear with the beam momentum. This dependence is shown in
figure~\ref{fig:ePlus_linearity_mc} for the $e^+$ data. Similar results (within
the errors)
are obtained for the $e^-$ data. 
 The lines indicate a fit with the
function \mbox{$\langle E_{\mathrm{vis}}\rangle =u + v\cdot p_{\mathrm{beam}}$}, where
$u$ is the offset, and $v$
the slope. The ratio between the
simulation and the data is shown in the bottom part of the figure. The
error bars indicate the overall uncertainties, i.e.\ the statistical
and systematic uncertainties added in quadrature.  
The data agree with the Monte Carlo simulations within uncertainties,
the deviations being less than 2\%.
The results of the linear fit are given in Table~\ref{tab:ePlus_linearity_mc_results}.
The offsets, which are consistent with zero,  are the combined
result of the 0.5~MIP threshold (loss of energy) and the detector
noise (addition of energy).

\begin{figure}[t!]
\begin{minipage}[c]{0.48\linewidth}
\centering
\includegraphics[scale=0.35]{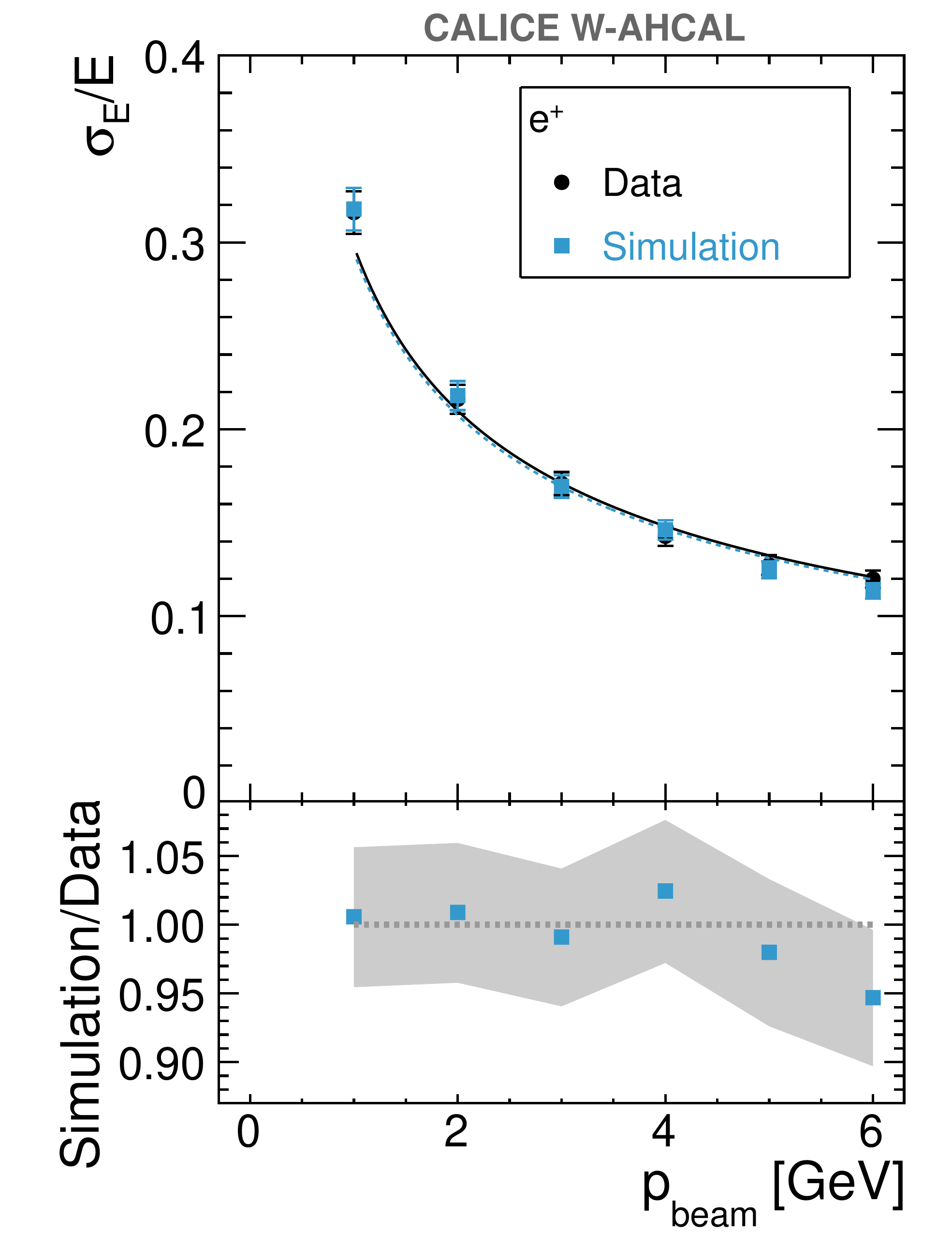}
\caption{Energy resolution for $e^+$ events: comparison of
  data with simulation. The error bars show the overall
  uncertainties. In the bottom part,
  the ratio between the simulation and the data is shown. The grey band shows the overall uncertainty.}
\label{fig:ePlus_resolution_mc}
\end{minipage}
\hspace{0.5cm}
\begin{minipage}[c]{0.48\linewidth}
\centering
    \captionof{table}{Parameters of the positron energy resolution fits for
      data and the simulation. The noise term is fixed to $1.06$~MIP.}
    \label{tab:ePlus_resolutionResults_mc}
    \begin{tabular}{lll}\toprule
      Parameter &  Data & Simulation \\\midrule
      $a$ [\%] &$29.6 \pm 0.5$  & $29.2 \pm 0.4$ \\
      $b$ [\%] & $0.0\pm 2.1$  &  $0.0 \pm 1.5$\\
       $c$ [GeV] & $0.036$ & $0.035$ \\
      $\chi^2$/ndf & 5.3/4 & 10.1/4\\\bottomrule
      \end{tabular}
\end{minipage}
\end{figure}

The $e^+$ energy resolution is presented in
figure~\ref{fig:ePlus_resolution_mc}. The fit function is:
\begin{equation}
\label{eq:resFit}
\displaystyle \frac{\sigma_E}{E}\equiv \frac{\sigma^{\mathrm{Novosibirsk}}}{\mathrm{\mu_{\mathrm{Novosibirsk}}}} =\frac{a}{\sqrt{E\;[\mathrm{GeV}]}}
\oplus b \oplus \frac{c}{E\;[\mathrm{GeV}]},
\end{equation}
where:
\begin{itemize}
\item $a$ is the \textbf{stochastic term}, which takes into account
  the statistical fluctuations in the shower detection.
\item $b$ is the \textbf{constant term}, which is dominated by
  the stability of the calibration, but includes also detector instabilities
  (i.e.\ non-uniformity of signal generation and collection, as well
  as loss of energy in
  dead materials);
\item $c$ is the \textbf{noise term}, the equivalent of the electronic
  noise in the detector, which includes noise from all the cells (with
  and without
  physical energy deposits). This term depends on the
  fiducial volume considered in the analysis.
\end{itemize}

The noise term $c$ is fixed to the spread (RMS) of the energy sum
distribution of randomly triggered noise events inside the beam
spill, considering only the central $3\times 3\;\mathrm{cm}^2$ tiles, 
contained in the first 20~layers, as done for the selection of the 
electromagnetic data (section~\ref{sect:emSelection}).
The measured noise RMS for the $e^+$ data is $(0.97\pm 0.01)$~MIP.
This value is converted into GeV
using the $v$ parameters of the fit given in
Table~\ref{tab:ePlus_linearity_mc_results}, resulting in 0.036~GeV.
The results of the fits to the 
$e^+$ energy spectra are shown in
table~\ref{tab:ePlus_resolutionResults_mc} for both data and
simulation. The results agree within the experimental uncertainties. 
A stochastic
term of \mbox{$(29.6 \pm 0.5)\%/\sqrt{E\;[\mathrm{GeV}]}$} is obtained for the CALICE
W-AHCAL, which is significantly higher than the stochastic term obtained for the
\mbox{CALICE} \mbox{Fe-AHCAL} of
\mbox{$(21.9 \pm 1.4)\%/\sqrt{E\;[\mathrm{GeV}]}$}
~\cite{{CALICE_AHCAL_emPaper}}. 
This degradation of the resolution is due to the coarser sampling of
the W-AHCAL with 2.8~$X_0$ per layer compared to 1.2~$X_0$ for the \mbox{Fe-AHCAL}.

\begin{figure}[t!]
\centering
\includegraphics[width=0.45\textwidth]{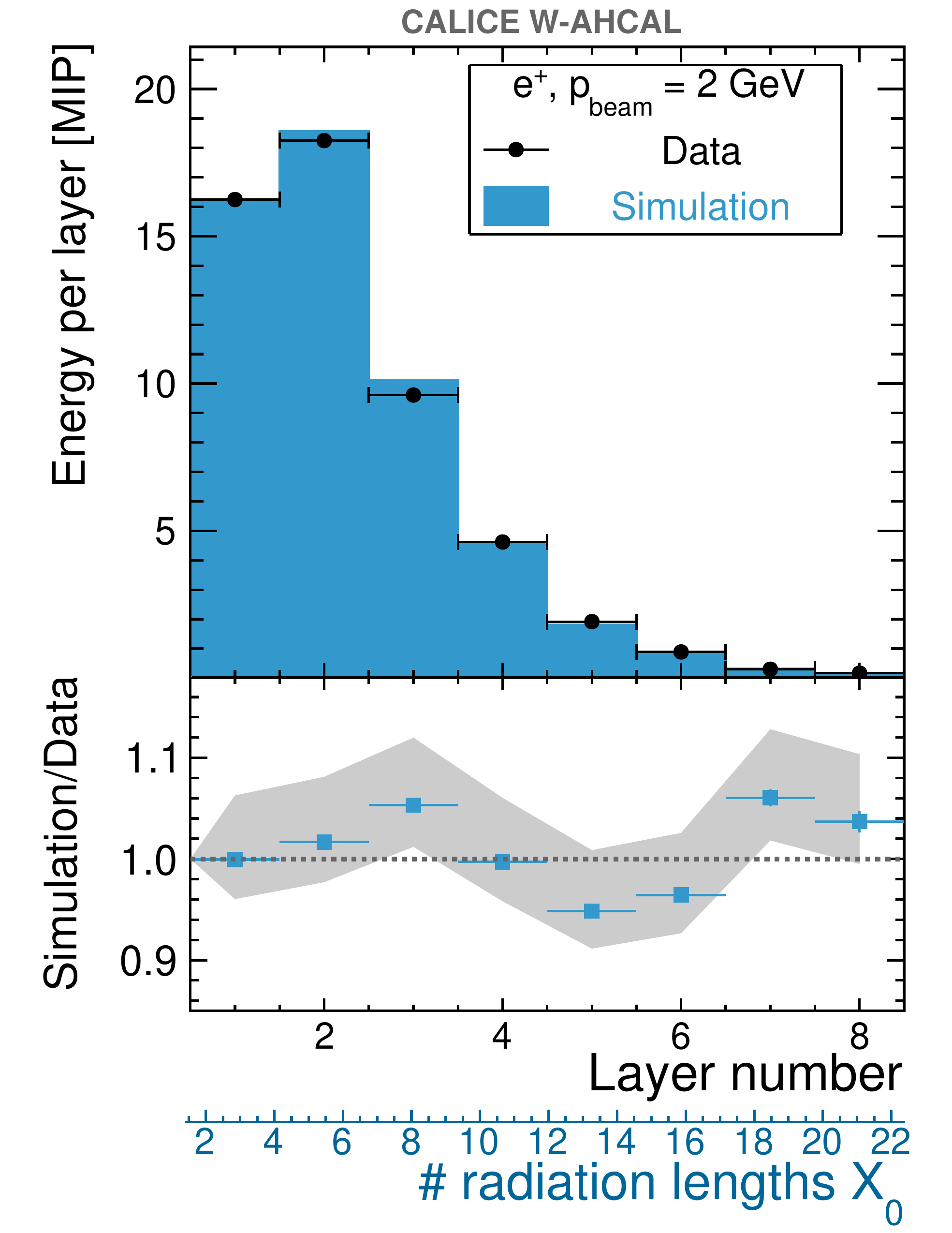}
\caption{Longitudinal shower profile for 2~GeV $e^+$:
  comparison of data with simulation. In the bottom part of the figure
  the ratio between the simulation and the data is shown. The grey band
  shows the overall uncertainty.}
\label{fig:ePlus_2GeV_longProfile}
\end{figure}

The longitudinal profile, i.e.\ the energy sum per layer as a function
of the calorimeter layer number, is shown in
figure~\ref{fig:ePlus_2GeV_longProfile} for 2~GeV~$e^+$. 
Due to the dense absorber material, most of the energy of the electromagnetic
shower is deposited in the first 5~calorimeter layers. The data and
the Monte Carlo simulation agree within the uncertainties, the deviations being
smaller than 10\% up to about 20~$X_0$. 

\section{Analysis of hadron data}
\label{sec:hadronAnalysis}

The selection of low energy hadrons is complicated by the presence
of muons from decays in flight, which are not sufficiently suppressed using Cherenkov
threshold counters. In addition, the energy sum distributions for
muons and pions overlap at low energies,
which makes the distinction more difficult. For this reason, only runs
with beam momenta from 3 (4) to 10~GeV/c are considered for the
$\pi^{\pm}$ (proton) analyses, as only for these was a
reliable selection of hadrons possible.

The pre-selection of hadron events is based on the Cherenkov threshold
counters.
In order to suppress the muons without the help of a tail catcher,
 information based on the
high granularity of the calorimeter is used. Algorithms are applied to identify
tracks~\cite{Weuste} and clusters~\cite{Lutz} in the calorimeters. A set of cuts on the number
of found tracks and on their length, as well as on the number of
clusters and their position in the calorimeter, was developed. It was
confirmed by comparison with the Monte Carlo simulation that the applied cuts
have no significant impact on the hadron events.

The events which fulfil any of the following cuts are considered to be either
muon-like
 or late showering hadrons:
\begin{itemize}
\item A track segment is identified which ends in layer $\geq 15$, has a small
  angle with respect to the normal incidence  ($\cos\phi\geq 0.99$), and traverses at least 14~layers;

\item At least two track segments are identified, which have a small
  angle ($\cos\phi>0.94$), each track traversing at least six layers;

\item At least one track segment is identified with hits in layer 29 or 30,
  and which traverses at least ten layers;

\item Two clusters are found in the first and second half of the
calorimeter, and they are aligned, i.e.\ the difference between their
$x$ and $y$ positions is less than the size of the scintillator tile
of 3~cm.
\end{itemize}
About 45\% of the events for hadrons with a beam momentum of 3~GeV/c
and about 50\% at 10~GeV/c fulfil the criteria described above and are
therefore rejected from the analysis.

\subsection{Analysis of pion data}
\label{sec:pionAnalysis}

The pre-selection of pions based on Cherenkov threshold counters
resulted in a sample with an electron and proton contamination of less
than~1\%. 

\begin{figure}[t!]
\centering
\includegraphics[width=0.45\textwidth]{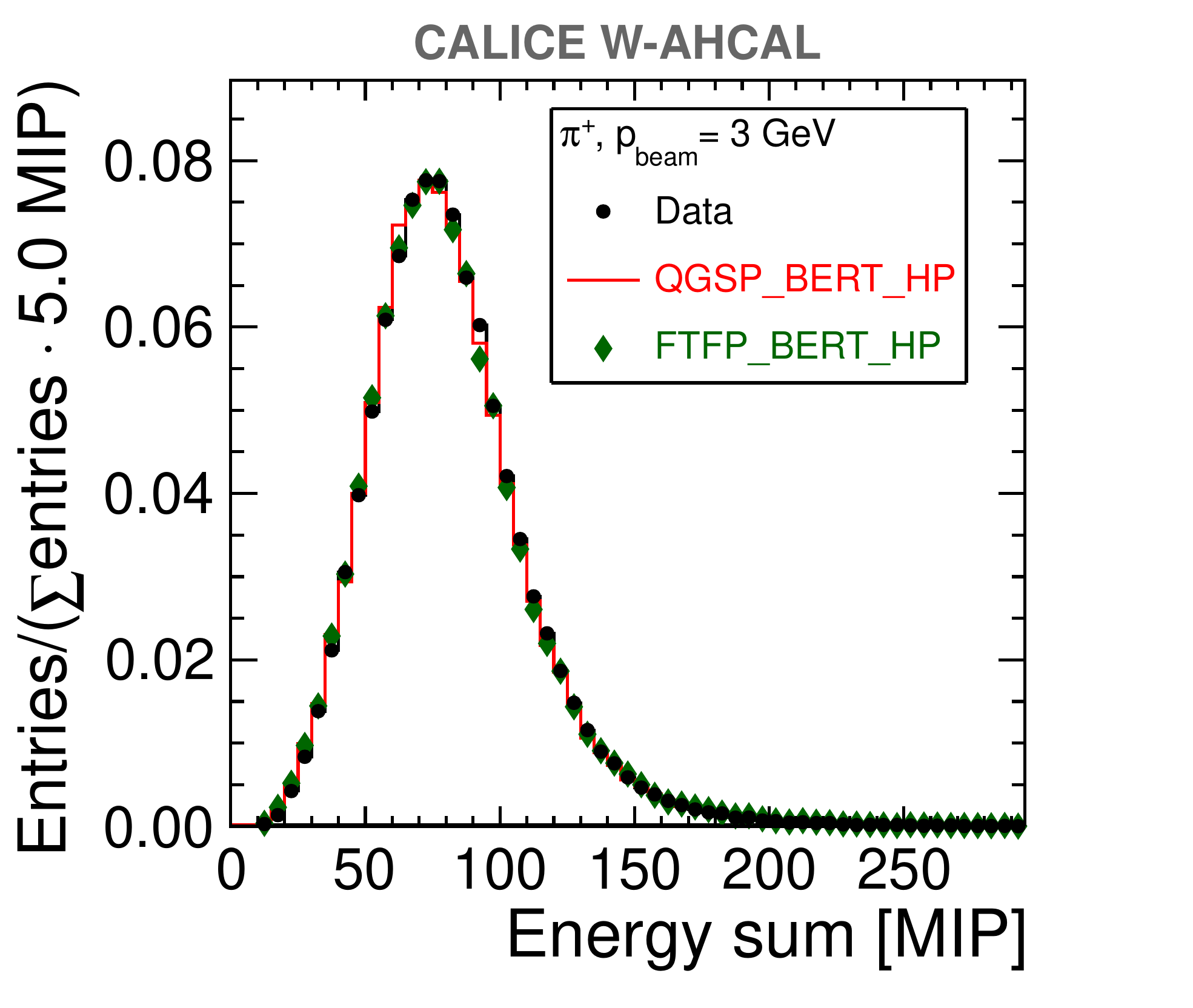}
\caption{Energy sum distribution for $\pi^+$ with a beam momentum of 3~GeV/c:
  comparison of data with selected GEANT4 physics lists.}
\label{fig:piPlus_3GeV_esum}
\end{figure}

 The hadron energy sum distributions are
non-Gaussian, with a high-energy tail, the effect being more
pronounced at low energies, as exemplified in
figure~\ref{fig:piPlus_3GeV_esum} for pions with a beam momentum of 3~GeV/c. This shape is predicted by the
selected GEANT4 physics lists.

\begin{figure}[t!]
\begin{minipage}[t]{0.48\linewidth}
\centering
\includegraphics[width=0.9\textwidth]{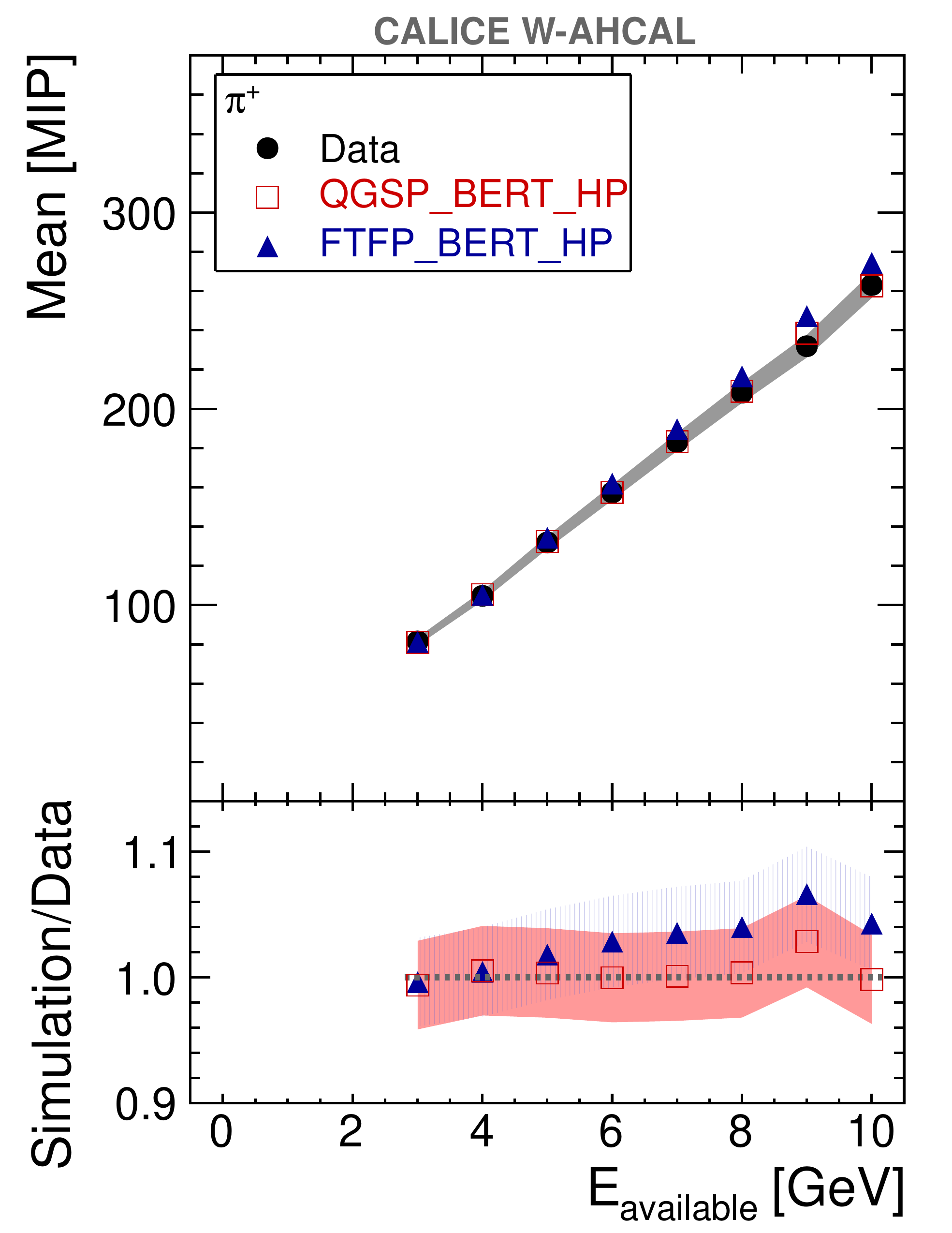}
\end{minipage}
\hspace{0.5cm}
\begin{minipage}[t]{0.48\linewidth}
\centering
\includegraphics[width=0.9\textwidth]{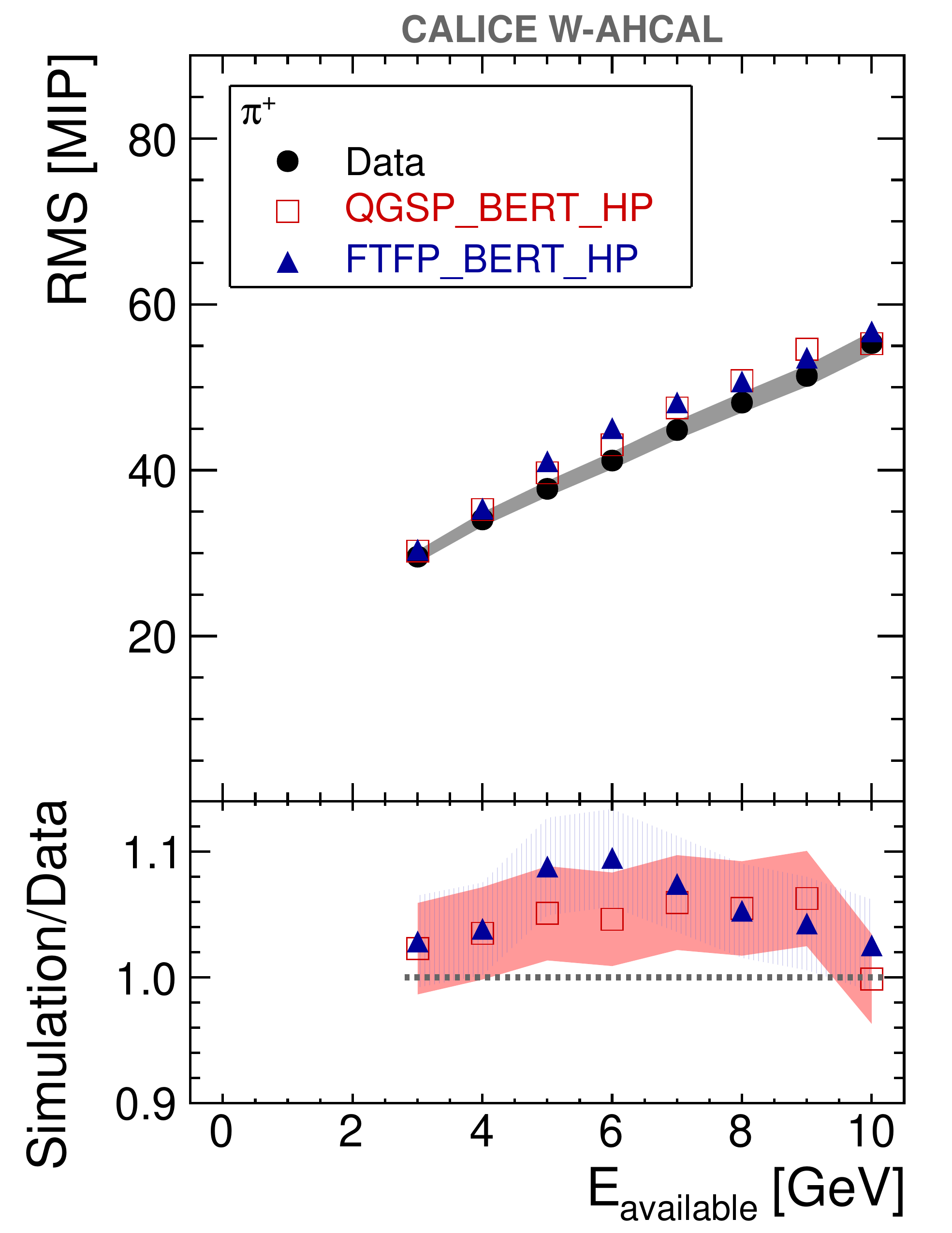}
\end{minipage}
\caption{Left: Dependence of the mean $\pi^+$ visible energy on the
  available energy. Right: Dependence of the RMS of the $\pi^+$ visible energy on the
  available energy.
   Data are compared with selected GEANT4 physics lists. In
  the bottom part of the figure the ratio between the simulation and
  the data
  is shown. The  bands
  show the overall uncertainty.}
\label{fig:piPlus_linearity_signa_mc}
\end{figure}

In order to measure the hadron energy resolution, we take the
non-Gaussian shape of the energy distributions into account  by using:
\begin{equation}
\label{eq:hadronResolution}
\displaystyle \frac{\sigma_E}{E}=\frac{\mathrm{RMS}}{\mathrm{Mean}},
\end{equation}
with RMS and
Mean obtained directly from the histogram
statistics.
The dependence of the mean visible energy on the available
energy $E_{\mathrm{available}}$
is shown in figure~\ref{fig:piPlus_linearity_signa_mc} (left), where $E_{\mathrm{available}}$
is the energy available for deposition in the calorimeter. In the case of
a pion, $E_{\mathrm{available}}$ is simply the particle's total energy~\cite{arxiv:1203.1302}:
\begin{equation}
\label{eq:piEavailable}
\displaystyle E_{\mathrm{available}} = \sqrt{p_{\mathrm{beam}}^2+ m_{\pi}^2},
\end{equation}
where $m_{\pi}=139.57$~MeV/c$^2$ is the pion mass. Data are compared with selected
GEANT4 physics lists. In the bottom part of the
figure~\ref{fig:piPlus_linearity_signa_mc} (left), the ratio between the simulation and
data is shown. 
The best description is given by QGSP\_BERT\_HP, the deviations being
of the order of 2\% or better.
As
FTFP\_BERT\_HP shares the same physics model for particles with
momenta up to 5~GeV/c, the agreement is
equally good, but gets worse when switching to the
Fritiof model. For both Bertini-based physics lists, a decrease of
the energy ratio is observed for 10~GeV/c. This corresponds to 
the transition to the Low Energy Parametrisation model for QGSP\_BERT\_HP. 
The RMS of the visible energy distributions is shown as a function of
the available energy in
figure~\ref{fig:piPlus_linearity_signa_mc} (right), for the different physics lists.
For  QGSP\_BERT\_HP and FTFP\_BERT\_HP the deviations are within 10\%.
The simulated distributions are in general broader than those of
the data.

\begin{figure}[t!]
\begin{minipage}[t]{0.48\linewidth}
\centering
    \includegraphics[width=0.9\textwidth]{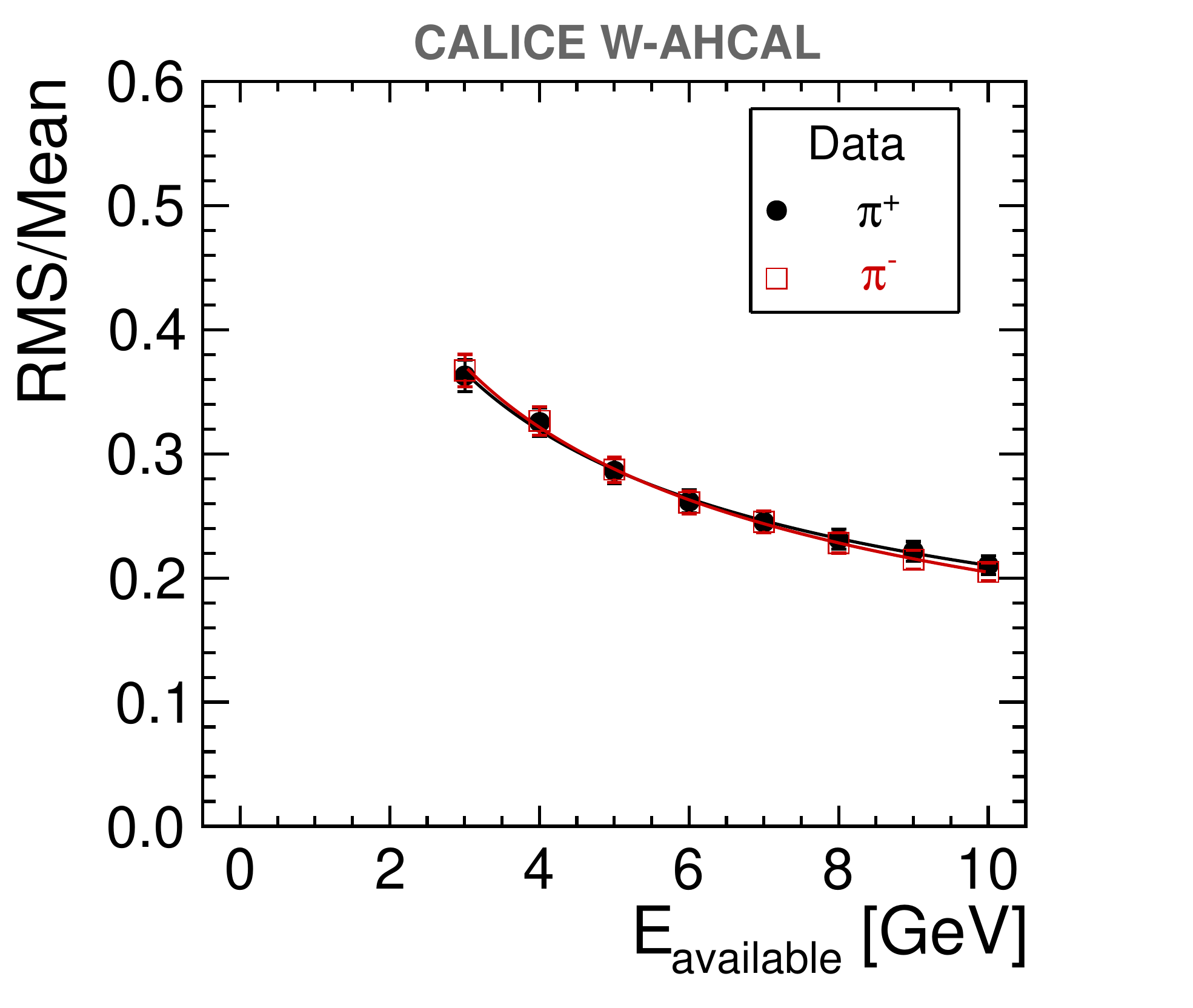}
    \caption{Energy resolution for the 2010 $\pi^\pm$ \mbox{W-AHCAL}
    data, obtained with equation~\protect\ref{eq:hadronResolution}. 
    The lines indicate a fit with the function given in equation~\protect\ref{eq:resFit}. 
    The error bars show the overall uncertainty.}    
    \label{fig:piPlusMinus_resolution}
\end{minipage}
\hspace{0.5cm}
\begin{minipage}[t]{0.48\linewidth}
\centering
\includegraphics[width=0.9\textwidth]{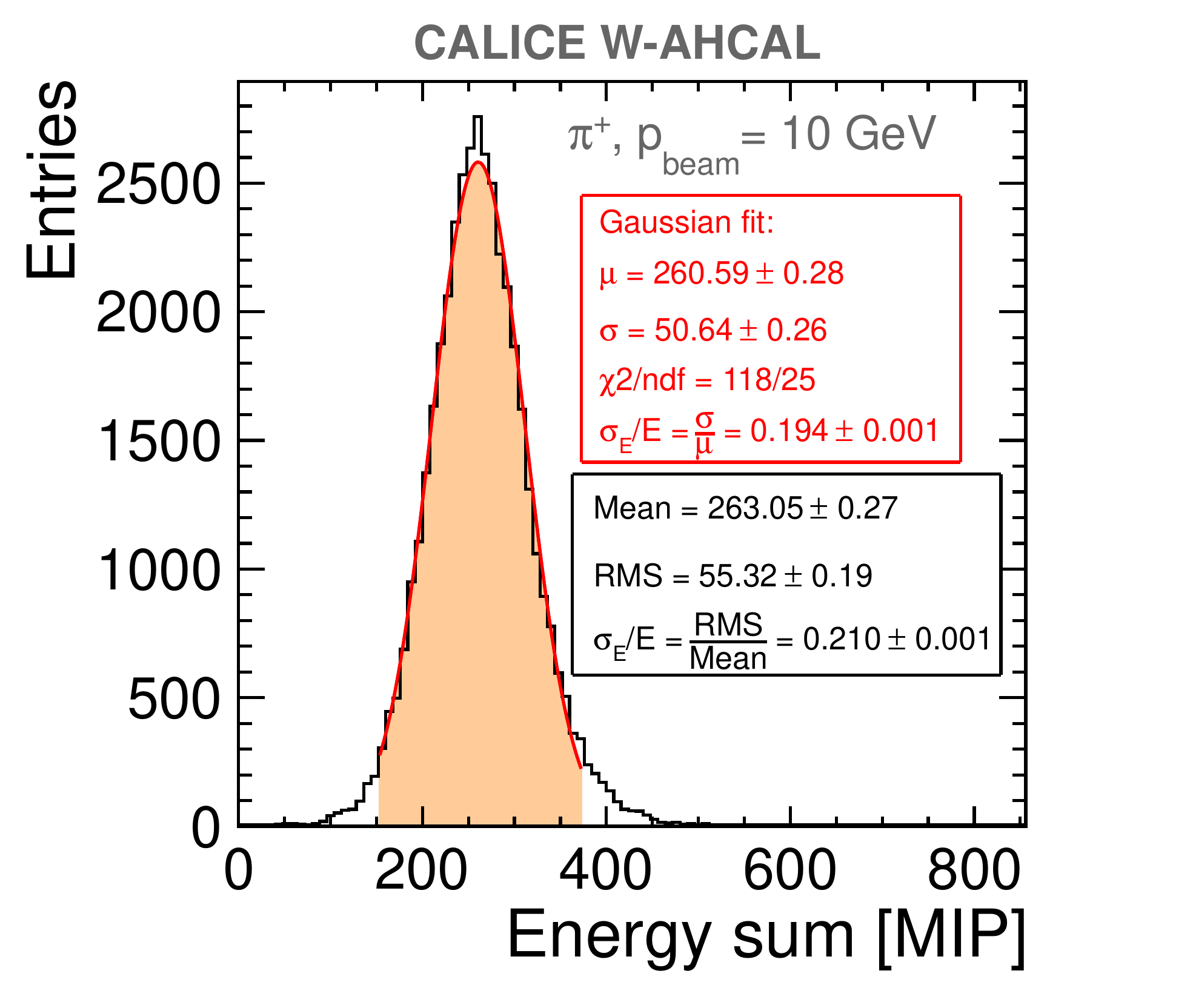}
\caption{Energy sum distribution for $\pi^+$ with a beam momentum of
    10~GeV/c. The red line indicates a fit with a Gaussian function in
    the region $\pm2\cdot$RMS around the mean (filled range). The energy resolutions
    obtained using the parameters of the Gaussian fit, as well as
    using the histogram statistics, are indicated.}
\label{fig:piPlus_10GeV_esum_comparisonOfEnergyResolutionMethods}
\end{minipage}
\end{figure}

The energy resolution for $\pi^{\pm}$ data is shown in
figure~\ref{fig:piPlusMinus_resolution}. The data are fitted with the
function defined in equation~\ref{eq:resFit}.
The $c$-term is fixed by the
spread~(RMS) of the energy distribution in randomly triggered events
inside the beam spill, considering all calorimeter cells. This term
amounts to 
71~MeV in the case of $\pi^-$ data, and to
70~MeV in the case of $\pi^+$ data.

The parameters obtained with the energy resolution fit
  are given in
 Table~\ref{tab:piPlusMinus_resolution_fit_results}. The stochastic
 term of \mbox{$(63.9 \pm 2.4)\%/\sqrt{E\;[\mathrm{GeV}]}$} is slightly worse than that
  of \mbox{$(57.6 \pm 0.4)\%/\sqrt{E\;[\mathrm{GeV}]}$} obtained for the CALICE
 Fe-AHCAL~\cite{CALICE_softwareCompensation}. However, a direct comparison of
  the pion resolutions measured with the two detectors is difficult
  due to several reasons. Firstly, in the W-AHCAL case the
  spectra have high energy tails, as illustrated in
 figure~\ref{fig:piPlus_10GeV_esum_comparisonOfEnergyResolutionMethods}. 
 Hence a Gaussian fit would result in a	too optimistic energy
  resolution, as indicated in the same figure. 
In the Fe-AHCAL case, the 
 energy spectra are fitted with a Gaussian function in a $\pm
 2\cdot$RMS range around the mean value. Secondly, the \mbox{Fe-AHCAL}
  data covered a much wider beam momentum range, from 10 to 100~GeV/c,
  compared to the range of 3 to 10~GeV in the W-AHCAL case. The $a$
  and $b$ parameters are anti-correlated, and poorly constrained with
  this low energy data, which is reflected in the large uncertainty of
  the $b$ parameter.

\begin{table}[t!]
\centering
    \caption{Parameters of the energy resolution fits for
          the 2010 \mbox{W-AHCAL} $\pi^\pm$ data. The $c$ parameter is
          fixed.}
    \label{tab:piPlusMinus_resolution_fit_results}
    \begin{tabular}{lll}\toprule
      Parameter                          &  $\pi^-$ & $\pi^+$ \\\midrule
      $a$ [\%] & $63.9\pm 2.4$ & $61.8\pm 2.5$\\
      $b$ [\%]                           & $3.2 \pm 6.9$ & $7.7\pm 3.0$\\
       $c$ [GeV]                         & $0.071$ &
          $0.070$ \\
      $\chi^2$/ndf                       & 0.4/6 & 0.5/6\\\bottomrule
      \end{tabular}
\end{table}

\begin{figure}[t!]
\begin{center}
{
\begin{minipage}[t]{0.9\linewidth}
\centering
\includegraphics[width=0.48\textwidth]{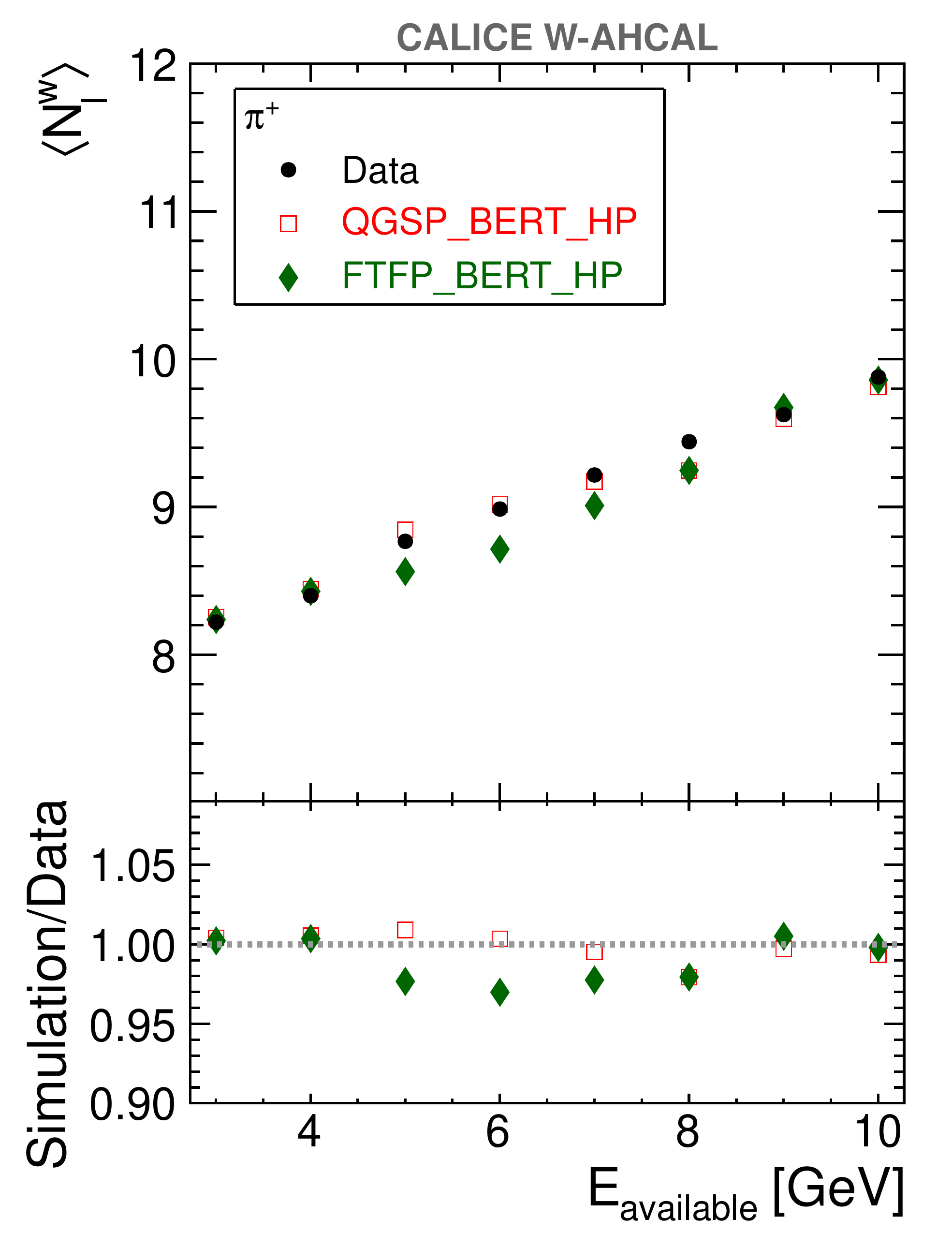}
\caption[]{Dependence of the mean energy weighted layer number of $\pi^+$
  initiated showers on the available energy:
  comparison of data with selected GEANT4 physics lists.  
   One layer corresponds to 0.13~$\lambda_{\mathrm{I}}$.
  In the bottom part of the figure the ratio
  between the simulation and the data is shown.}
\label{fig:piPlus_meanKcog_qgsp_bert_hp}
\end{minipage}
}
\end{center}

\begin{minipage}[t]{0.48\linewidth}
\centering
\includegraphics[width=0.9\textwidth]{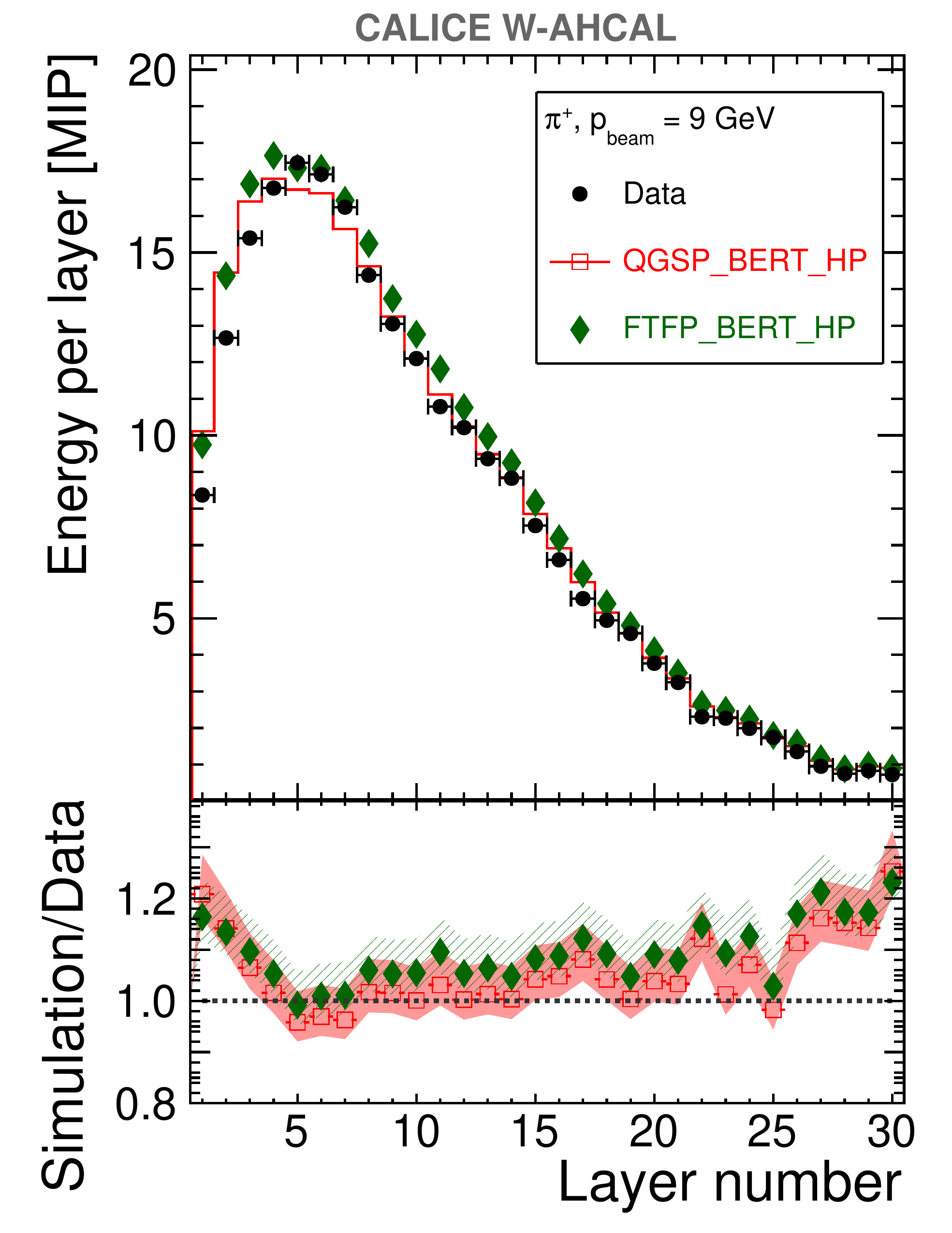}
\end{minipage}
\hspace{0.5cm}
\begin{minipage}[t]{0.48\linewidth}
\centering
\includegraphics[width=0.9\textwidth]{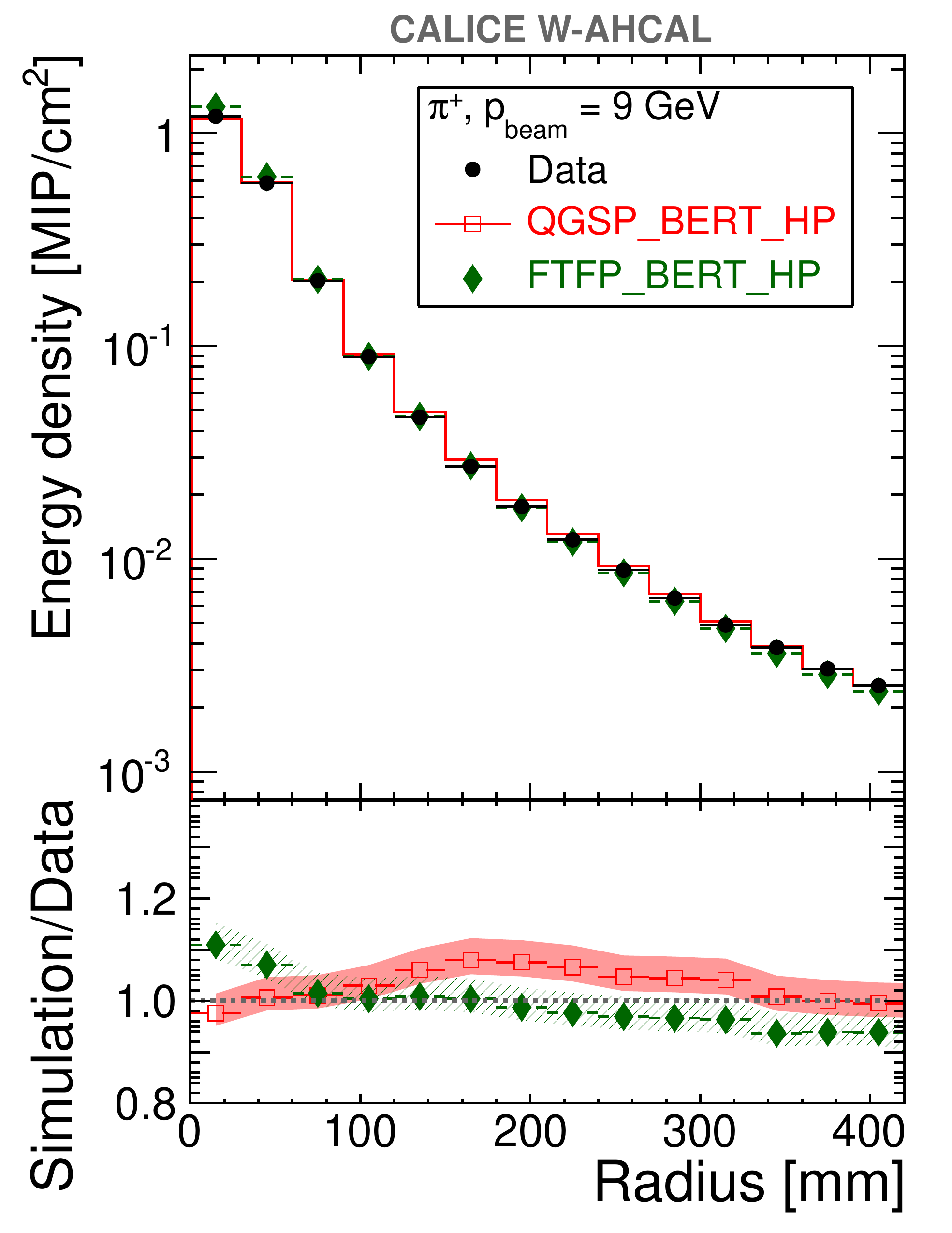}
\end{minipage}
\caption{Longitudinal (left) and radial  (right) shower profiles of $\pi^+$ with a beam momentum of 9~GeV/c:
  comparison of data with selected GEANT4 physics lists. In the bottom part of the
  figures the ratios between the simulations and the data are displayed. The  bands
  show the overall uncertainty.}
\label{fig:piPlus_9GeV_longProfile_radialProfile}
\end{figure}

In order to judge the quality of the simulation concerning the spatial
development of hadron showers, comparisons of data with the Monte
Carlo simulation
were done for variables which describe the shower development
along the $z$-axis (longitudinally) and in the ($x,\; y$) plane (transversely).
To study the longitudinal shower development, a variable called
the energy weighted layer number is defined as:
\begin{equation}
\langle N_l^{w} \rangle =  \displaystyle \frac{\sum_i E_i \cdot \mathrm{layer}_i}{\sum_i E_i}
\end{equation}
where $E_i$ is the hit energy in cell $i$, layer$_i$ is the layer
number to which cell $i$ belongs, and the summation is done over all
cells.
This variable is sensitive to the longitudinal shower development: the
mean energy weighted layer $\langle N_l^w \rangle$ will have a larger value
for showers which develop deep in the calorimeter than for early
starting showers.
The dependence of the mean energy weighted layer number on
the $\pi^+$ available energy is presented in
figure~\ref{fig:piPlus_meanKcog_qgsp_bert_hp}, which contains also 
the ratio between the simulation and the data.
 The observed disagreement is within $\pm 3\%$ for both QGSP\_BERT\_HP and
 FTFP\_BERT\_HP.

\afterpage{\clearpage}

%

The longitudinal profile for $\pi^+$ with a beam momentum of 9~GeV/c is
shown in figure~\ref{fig:piPlus_9GeV_longProfile_radialProfile}
(left).
In the central part the energy deposition is well reproduced by the
simulation models considered. However, both models overestimate the
energy depositions in the first and last calorimeter layers
by up to 25\%. The difference in the front part of the calorimeter
cannot be related to an improper description of the material in the
test beam, since the longitudinal profile of 2~GeV~$e^+$ is well
described as shown in figure~\ref{fig:ePlus_2GeV_longProfile}.
The simulation models seem instead to predict an earlier shower start
than observed in the experimental data.

The shower development in the transverse plane is studied by means of
the so-called radial profile. The procedure to measure this profile is
the following: In order to reduce the influence of the varying detector granularity
within one layer, the physical W-AHCAL
cells are divided into virtual cells of $1\times
1$~cm$^2$~\cite{Lutz}. In a next step, the energy in a given cell is distributed
randomly among the $1\times1$~cm$^2$ virtual cells contained in the
real cell. Then virtual rings, centred on ($x_{\mathrm{cog}},\; y_{\mathrm{cog}}$),
are built. The radii of these rings are multiples of the width of the
smallest W-AHCAL tile, i.e.\ 3~cm. 
Next the energy density, i.e.\ the energy contained in a given ring
divided by the area of the ring, is measured in MIP/cm$^2$ in each ring.
Finally, the radial profile is given by the
distribution of the energy density (i.e., energy per unit area) as a
function of the radial distance to the shower centre-of-gravity, defined as:
\begin{equation}
\label{eq:radial_distance}
r_i = \displaystyle\sqrt{(x_i-x_{\mathrm{cog}})^2 + (y_i-y_{\mathrm{cog}})^2},
\end{equation}
where $x_i$ ($y_i$) is the $x$ ($y$) position of the centre of the cell $i$, 
and $x_{\mathrm{cog}}$ and $y_{\mathrm{cog}}$ are the centres-of-gravity
in $x$ and $y$ for the whole calorimeter: 
\begin{equation}
x_{\mathrm{cog}}=\displaystyle \frac{\sum_i E_i \cdot x_i}{\sum_i
  E_i} \; \mathrm{and} \; y_{\mathrm{cog}}=\displaystyle
  \frac{\sum_i E_i \cdot y_i}{\sum_i E_i},
\end{equation}
with $E_i$ being the hit energy in cell $i$.

An example of a radial profile is given in
figure~\ref{fig:piPlus_9GeV_longProfile_radialProfile} (right) for
$\pi^+$ with a beam momentum of 9~GeV/c. 
The deviations are at the level of 10\% or smaller for QGSP\_BERT\_HP,
which describes the data better than FTFP\_BERT\_HP.


\subsection{Analysis of proton data}
\label{sec:protonAnalysis}
The calorimeter response to protons differs from the response to pions mainly
due to two effects~\cite{CDF}:
\begin{itemize}
\item The first effect is due to the differences in the 
  energy available for deposition in the calorimeter.
  For pions, it is given in equation~\ref{eq:piEavailable}. 
  For protons, the available energy is:
  \begin{equation}
    \label{eq:protonEavailable}
    \displaystyle E_{\mathrm{available}} = E_{\mathrm{kin}} =
  \sqrt{p_{\mathrm{beam}}^2+ m_{\mathrm{proton}}^2} -
  m_{\mathrm{proton}},
  \end{equation}
  where $m_{\mathrm{proton}}=938.27$~MeV/c$^2$ is the proton mass.
  This is relevant for the
  low energy range analysed in this paper;

\item The second effect originates from the different fractions of $\pi^0$
  mesons produced in proton and pion-induced showers. As a
  consequence of baryon number conservation, which favours the
  production of leading baryons, one expects a smaller average number of
  $\pi^0$ mesons in proton showers, compared to pion showers. In the
  latter case, the leading
  particle may be a $\pi^0$, due to the charge exchange
  reaction:
  $\pi^+  n   \rightarrow \pi^0 + p$. This reaction is favoured by the large number of
  neutrons in tungsten, i.e.\ about 50\% more neutrons than protons.
 A smaller number of $\pi^0$
  implies a smaller electromagnetic fraction in the shower. For a
  non-compensating calorimeter (\mbox{$e$/$h>1$}), this results in a
  higher response for pions than for protons.
\end{itemize}

The selection cuts for protons are the same as for pions, apart from
the Cherenkov-based particle identification. 
Only data with beam momenta from 4 to 10 GeV/c are included for the proton
analysis.
 For this momentum range, electrons and pions are rejected with
high efficiency based on the signals from the Cherenkov threshold
counters~\cite{LCD-Open-2013-001}, 
resulting in samples with
negligible $e^+$ and $\pi^+$ contamination (less than 1\%). 
The remaining muons were rejected as described in section~\ref{sec:pionAnalysis}.
The procedure to measure the energy and resolution for protons is
the same as for pions.

\begin{figure}[t!]
\begin{minipage}[t]{0.48\linewidth}
\centering
\includegraphics[width=0.9\textwidth]{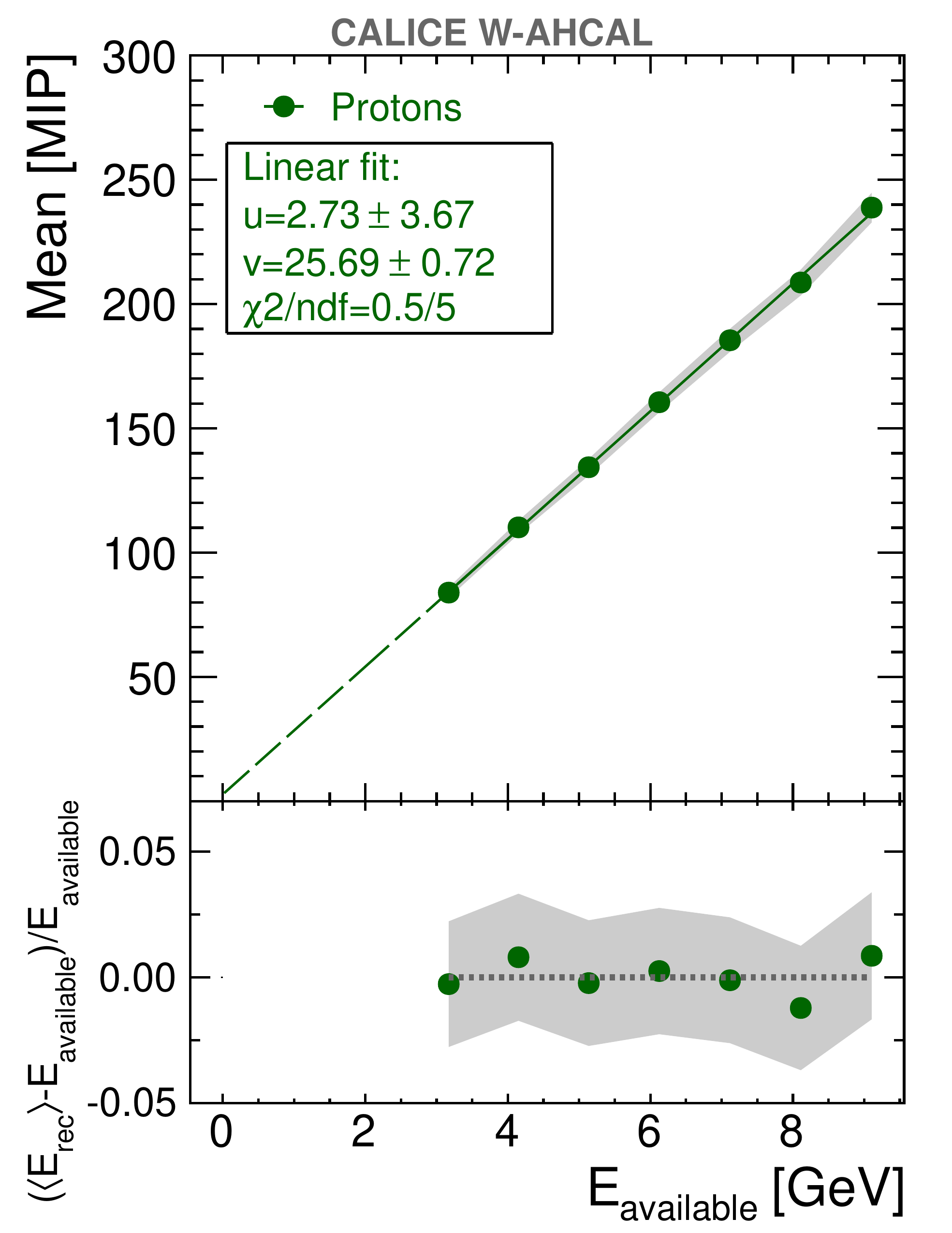}
\caption{Dependence of the mean visible proton energy Mean on the
  available energy.  The
  line indicates a fit with the
  function \mbox{$\textrm{Mean} =u + v\cdot
  E_{\mathrm{available}}$}. The fit parameters are also indicated. 
  In the bottom part of the figure, the residuals to the fit are
  displayed, with $\langle
  E_{\mathrm{rec}}\rangle\;\mathrm{[GeV]}=(\mathrm{Mean}\; \mathrm{[MIP]}-u)/v.$
  The grey band shows the overall uncertainty.}
\label{pLinearityData}
\end{minipage}
\hspace{0.5cm}
\begin{minipage}[t]{0.48\linewidth}
\centering
\includegraphics[width=0.9\textwidth]{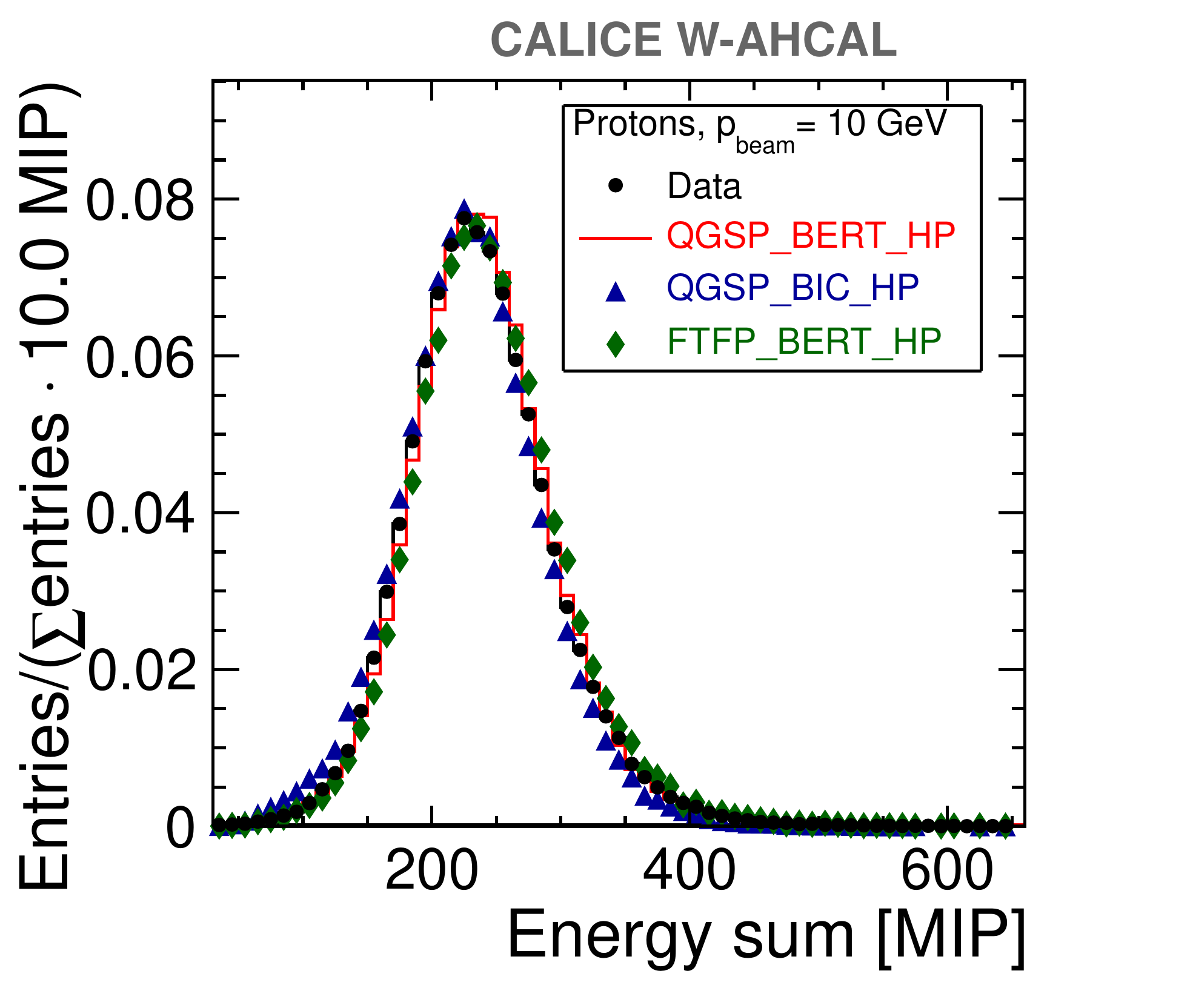}
\caption{The visible energy distribution of protons with a beam
  momentum of 10~GeV/c: comparison of data with selected GEANT4
  physics lists.}
\label{fig:proton_esum_10GeV}
\end{minipage}
\end{figure}

The average calorimeter response for protons is shown as a function of
the available beam
energy  in figure~\ref{pLinearityData}. 
The residuals to a
linear fit are shown in the bottom part of the same figure.
The proton response is linear within
the experimental uncertainties.
The proton visible energy distribution is compared with the
expectation from selected GEANT4 physics lists in
figure~\ref{fig:proton_esum_10GeV} for the 10~GeV/c
case. The level of agreement between data and the simulation models is
very good.

\begin{figure}[t!]
\begin{minipage}[t]{0.48\linewidth}
\centering
\includegraphics[width=0.9\textwidth]{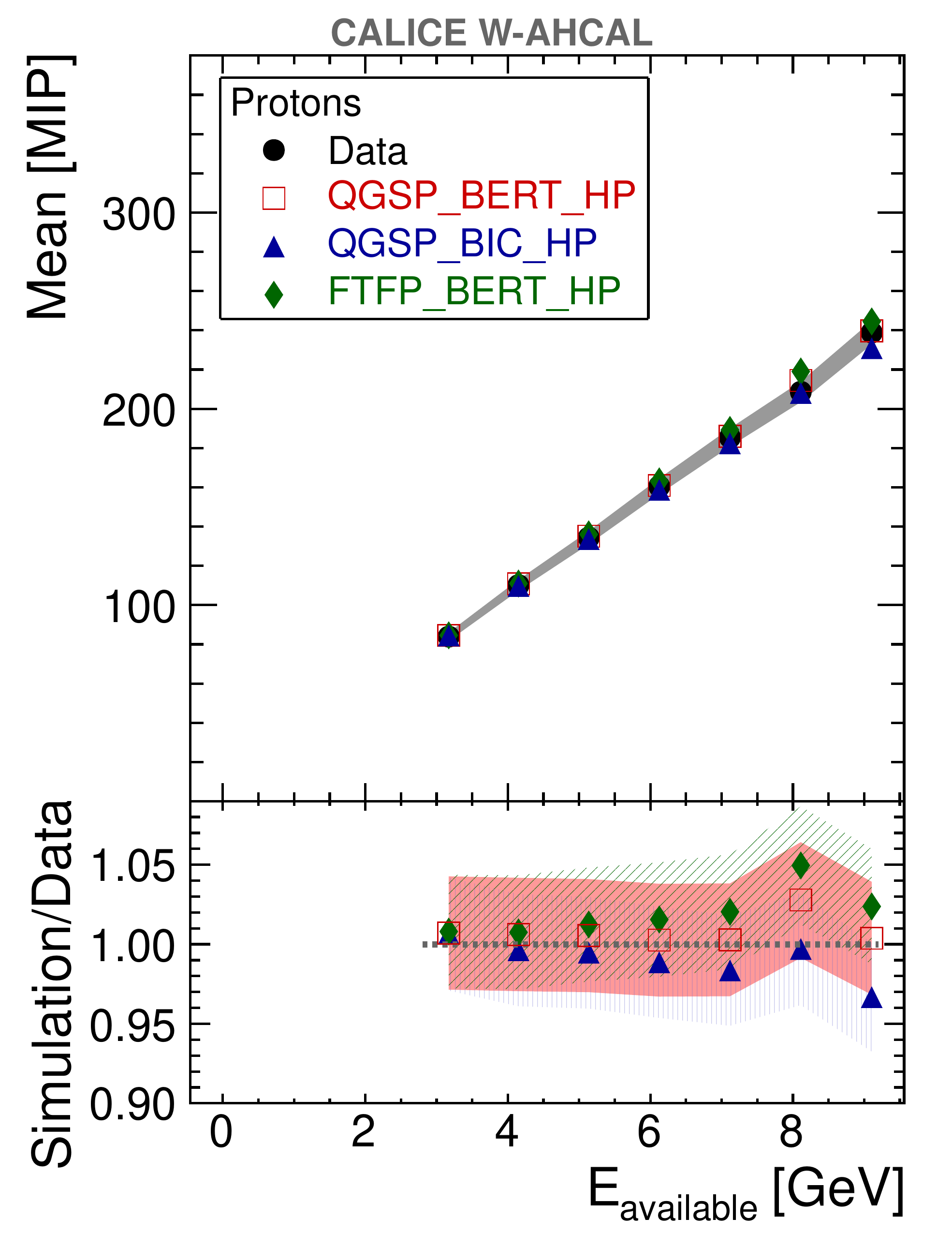}
\end{minipage}
\hspace{0.5cm}
\begin{minipage}[t]{0.48\linewidth}
\centering
\includegraphics[width=0.9\textwidth]{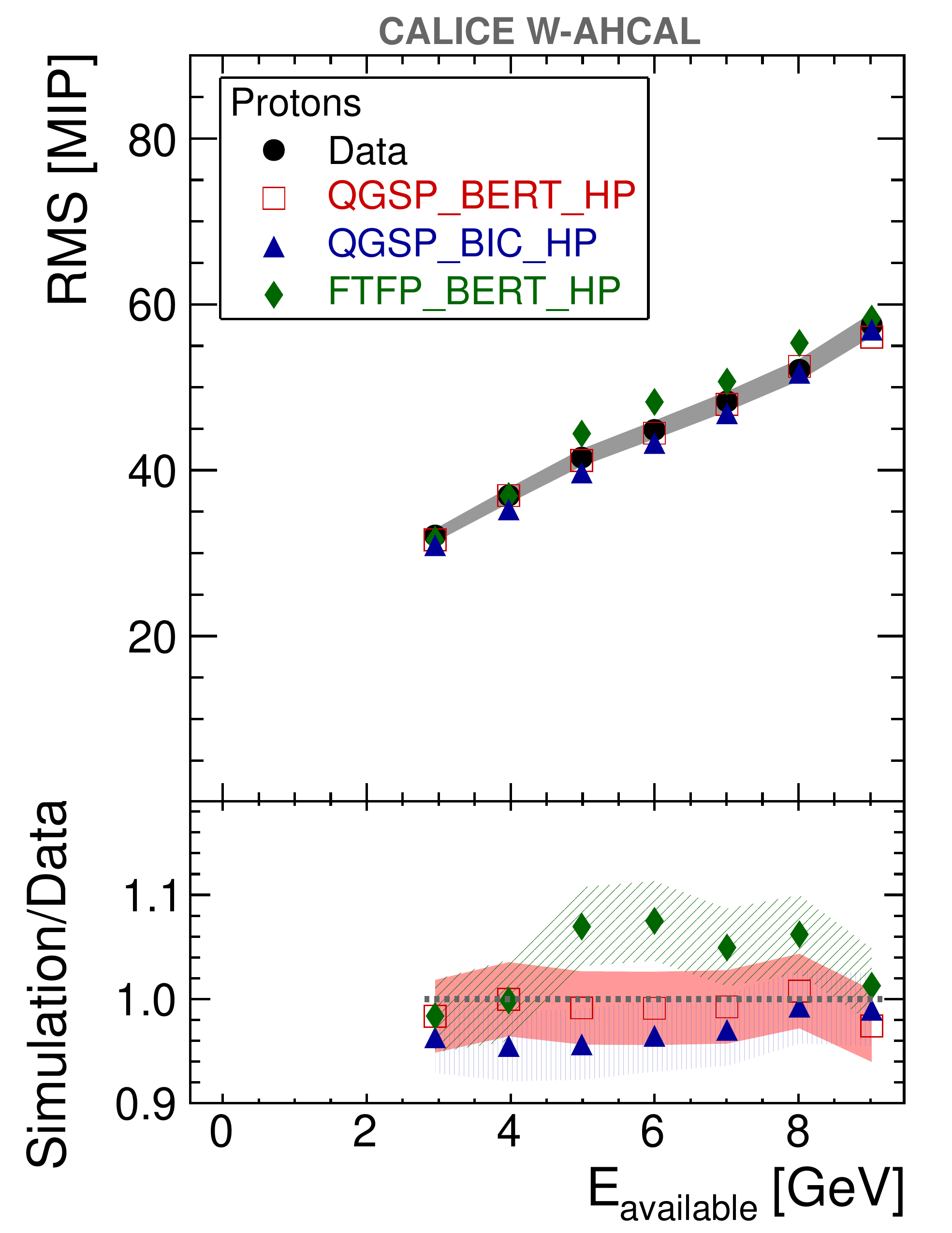}
\end{minipage}
\caption{Left: Dependence of the mean visible proton energy $\langle
  E_{\mathrm{vis}}\rangle$  on the
  available energy. Right: Dependence of the RMS of the proton energy distributions on the
  available energy.
  The error bands show the overall uncertainty. Data
  are compared with selected GEANT4
  physics lists. In the bottom part of the figures the ratios between
  the different simulation models and the data are shown. }
\label{fig:proton_linearity_sigma_mc}
\end{figure}

The proton mean visible energy as a function of the available energy is
compared for data and the selected GEANT4 physics lists 
in figure~\ref{fig:proton_linearity_sigma_mc} (left). 
As in the pion case,
the best description is given by
 the QGSP\_BERT\_HP physics list, the differences
being less than~2\%. For protons, QGSP\_BIC\_HP also performs well,
although the agreement becomes worse with increasing available energy.
The RMS of the
energy distribution is displayed as a function of the available energy
in figure~\ref{fig:proton_linearity_sigma_mc} (right). Several steps are observed in
the ratio between simulation and the data, corresponding to the transition
from one simulation model to another. For example in the FTFP\_BERT\_HP
physics list, the transition  from the
Bertini cascade to the FTFP model is between 4 and
5~GeV. However, in all cases the deviations between simulation and
data are smaller than 10\%.

\begin{figure}[t!]
\begin{minipage}[t]{0.48\linewidth}
\centering
\includegraphics[scale=0.35]{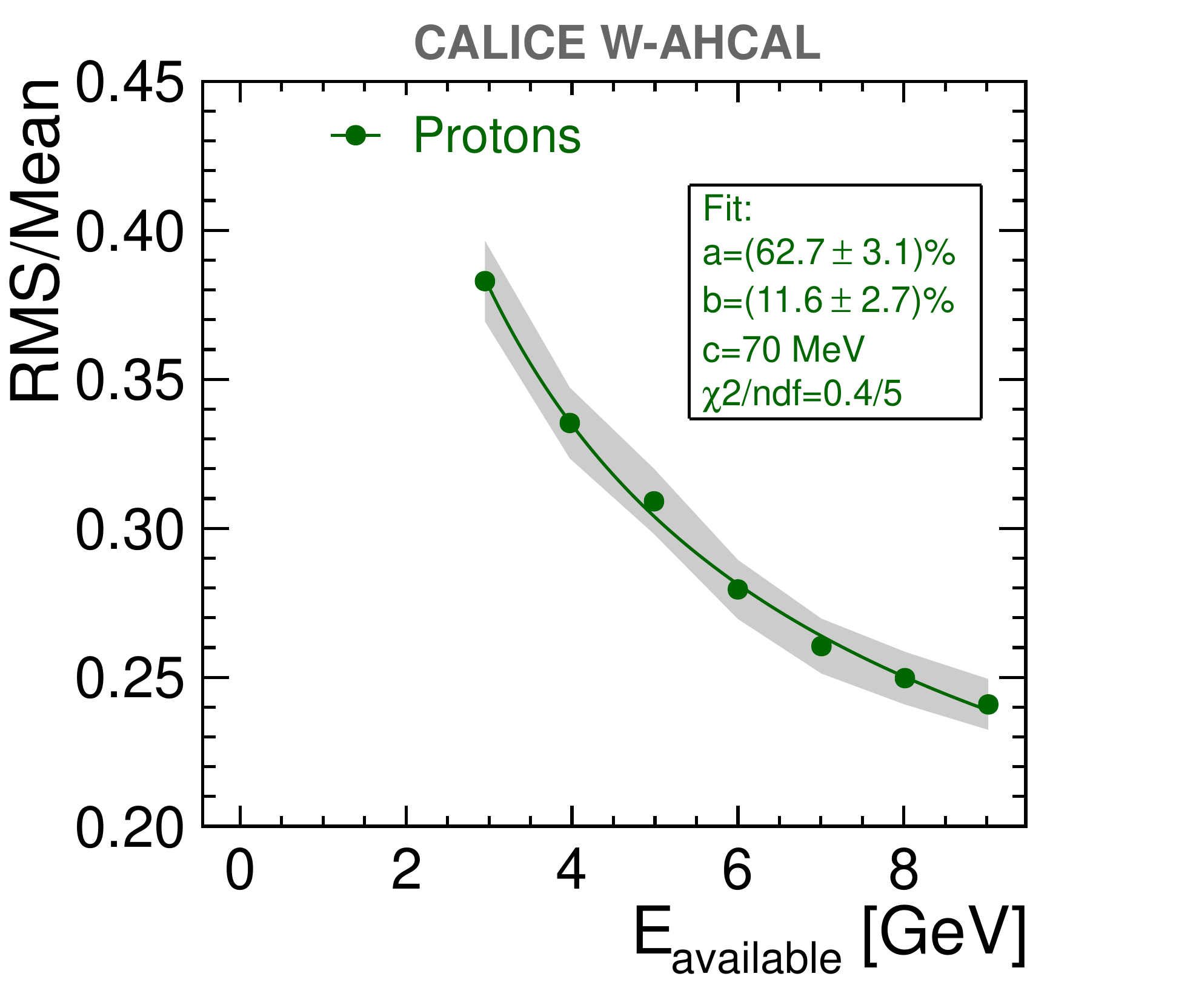}
\caption{Proton energy resolution defined as RMS/Mean, as a function
  of the available energy. The
  grey band shows the overall uncertainty. The fit parameters are also indicated. }
\label{fig:proton_energyResolution}
\end{minipage}
\hspace{0.5cm}
\begin{minipage}[t]{0.48\linewidth}
\centering
\includegraphics[scale=0.35]{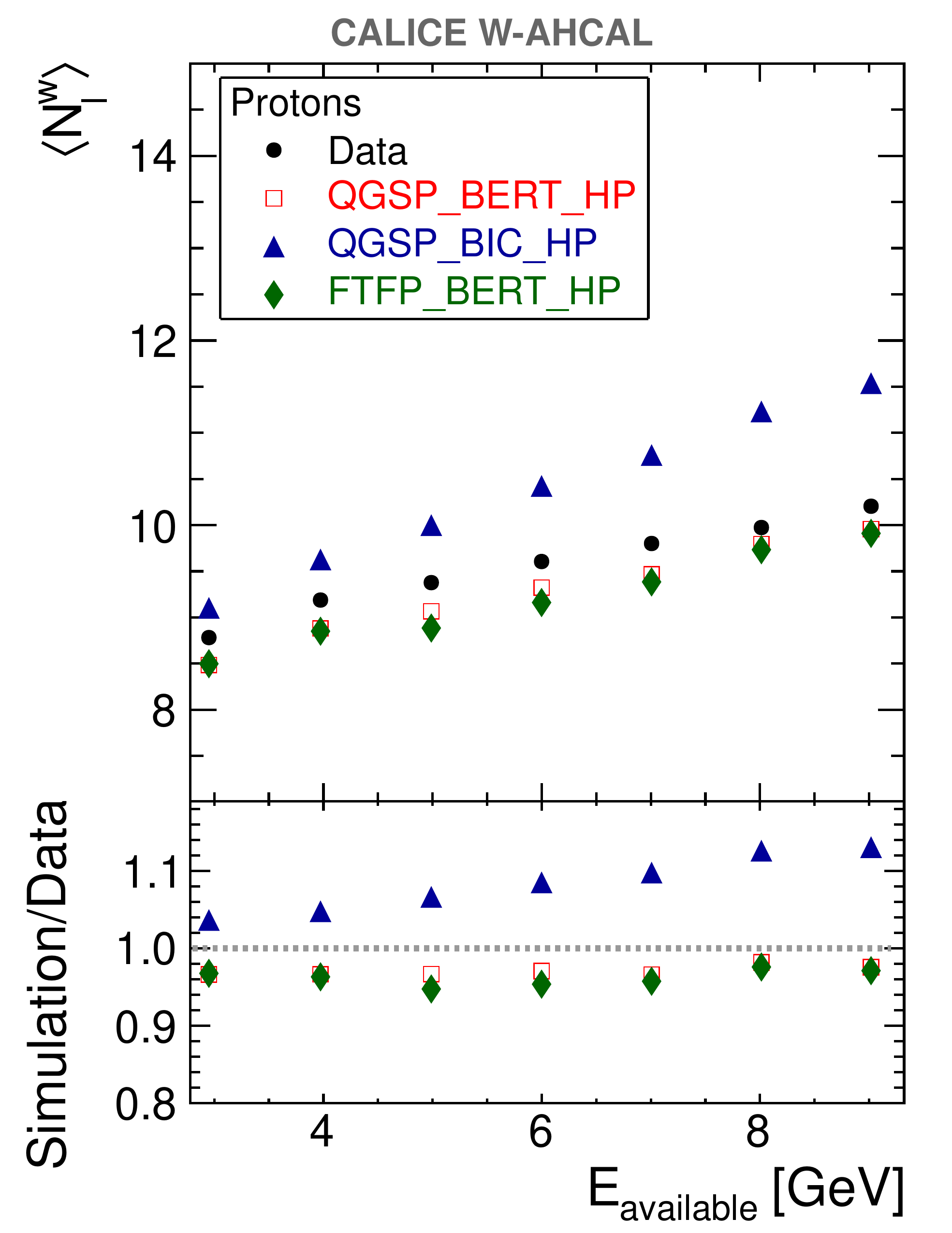}
\caption[]{Dependence of the mean energy weighted layer number  for proton initiated showers vs.\ the
available energy: comparison of data with selected GEANT4 physics lists. One layer
corresponds to 0.13~$\lambda_{\mathrm{I}}$.
  In the bottom part of the figure the ratio
  between the simulation and the data is shown.}
\label{fig:proton_kcog_qgsp_bert_hp}
\end{minipage}
\end{figure}

The
proton energy resolution, obtained using
equation~\ref{eq:hadronResolution}, is presented in
figure~\ref{fig:proton_energyResolution}. 
The parameters of the fit with the function given by
equation~\ref{eq:resFit} are also displayed.
The noise term is fixed
to the same value of 70~MeV as for the $\pi^+$ data.
The stochastic term of
\mbox{$(62.7 \pm 3.1)\%$/$\sqrt{E\;[\mathrm{GeV}]}$} is comparable with the
value obtained in the
$\pi^+$ case, \mbox{$(61.8\pm 2.5)\%$/$\sqrt{E\;[\mathrm{GeV}]}$}.
The main difference is the constant term, which is higher:
\mbox{$(11.6 \pm 2.7)\%$} for protons, compared to \mbox{$(7.7 \pm 3.0)\%$} for pions. This is compatible
with expectations from simulations. QGSP\_BERT\_HP 
predicts a stochastic term of about 62\%, and a constant term of about
\mbox{$11\%$}. For a better constraint on the constant terms, it would be
necessary to also include higher energy data in the fit.

\begin{figure}[t!]
\begin{minipage}[t]{0.48\linewidth}
\centering
\includegraphics[width=0.9\textwidth]{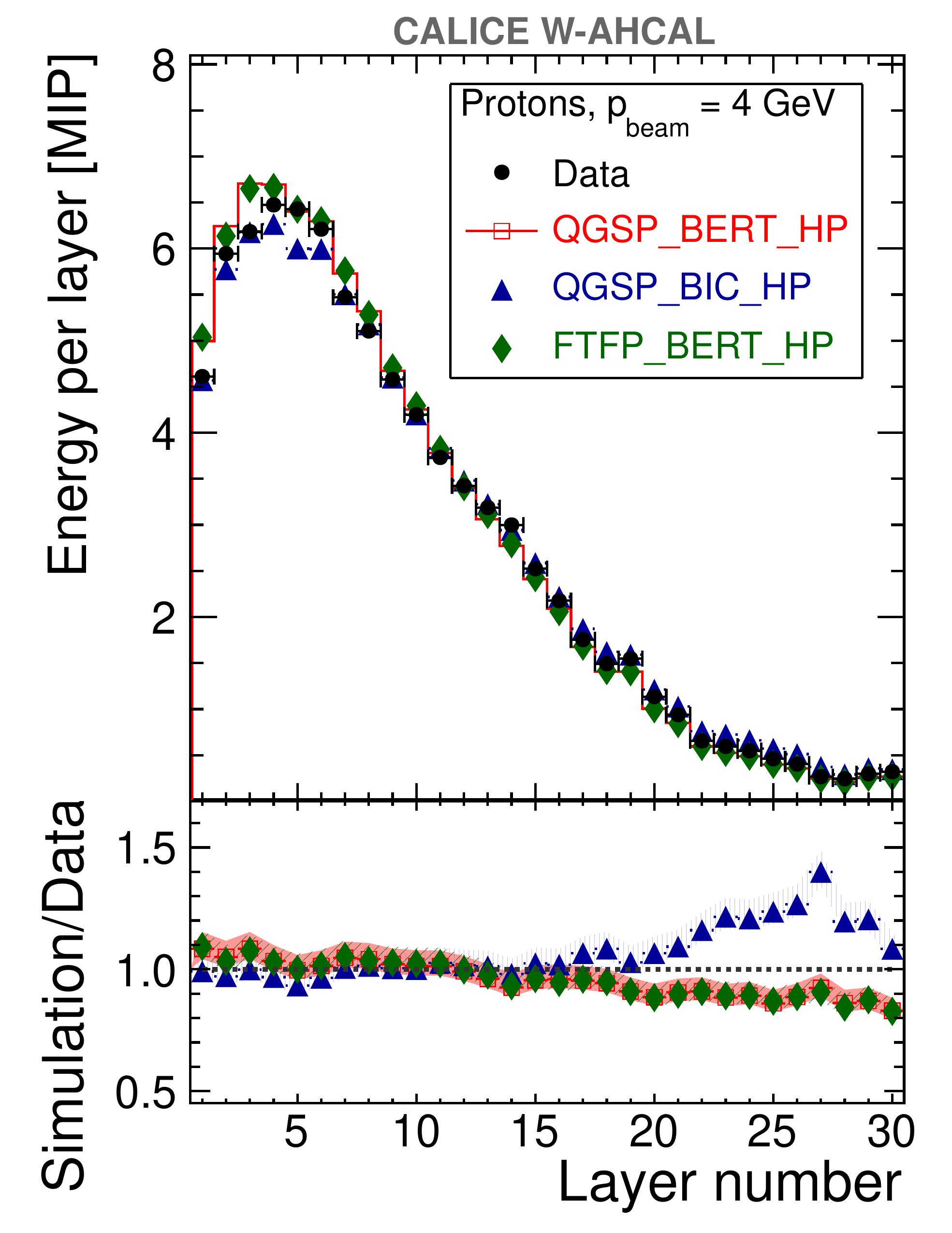}
\end{minipage}
\hspace{0.5cm}
\begin{minipage}[t]{0.48\linewidth}
\centering
\includegraphics[width=0.9\textwidth]{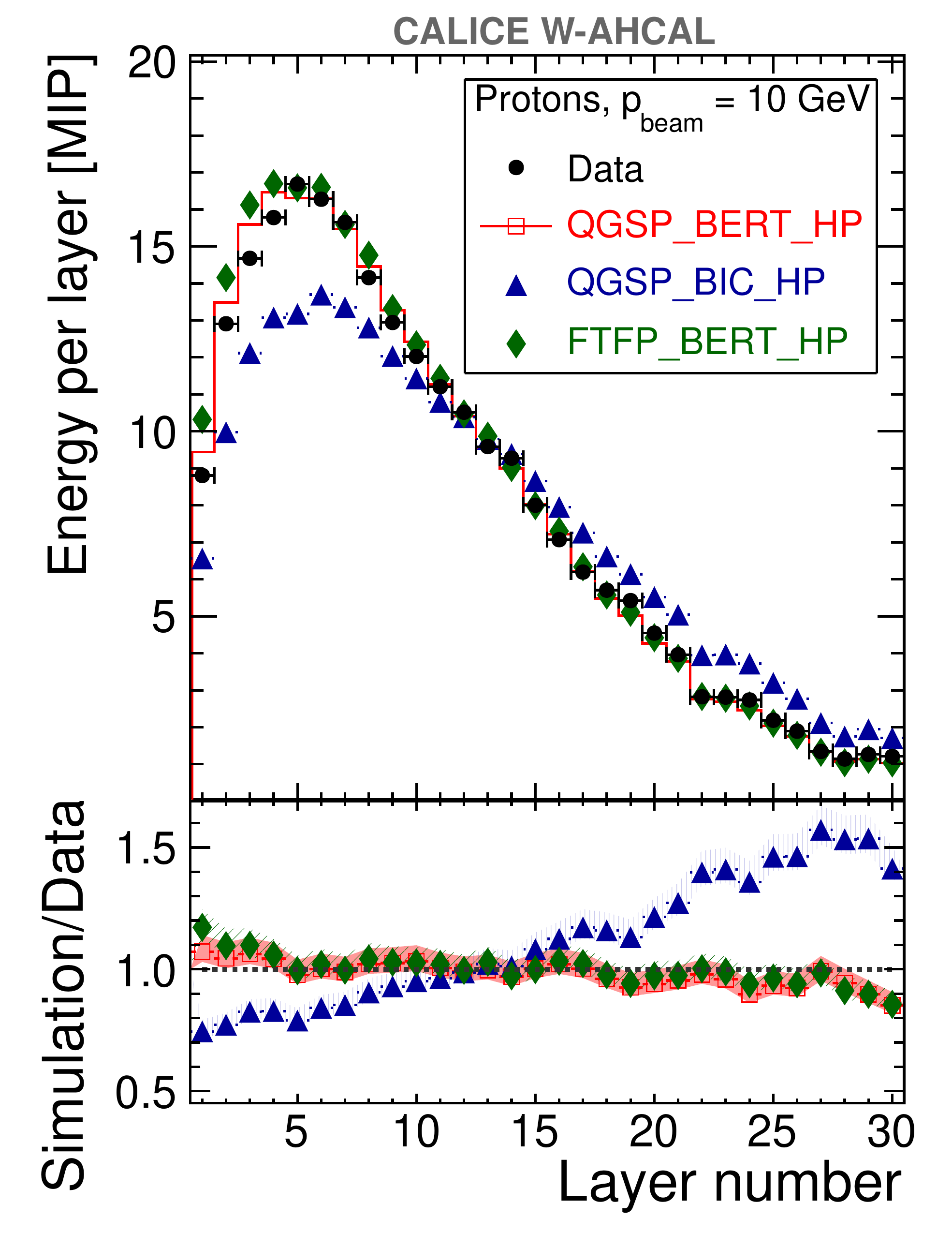}
\end{minipage}
\caption{Longitudinal shower profile for a proton with a beam momentum
  of 4~GeV/c (left) and of 10~GeV/c (right). Data are compared with selected
  GEANT4 physics lists. In the bottom part of the figures the ratios
  between the simulations and the data are displayed. The  bands
  show the overall uncertainties.}
\label{fig:proton_4_10GeV_longProfile}
\end{figure}

The dependence of the mean energy weighted layer number on the available energy is presented
in figure~\ref{fig:proton_kcog_qgsp_bert_hp}, together with the ratios
of selected GEANT4 physics lists to the data. 
The Bertini-based models (QGSP\_BERT\_HP and FTFP\_BERT\_HP) show the
best agreement with data, the deviations being less than 3\%.
QGSP\_BIC\_HP, on the other side, predicts higher values
than observed, i.e.\ the showers start to develop later in the
calorimeter,
 and the differences are increasing with the available energy.
This behaviour can also be observed in 
the longitudinal shower profiles for protons with beam momenta of 4 and
10~GeV/c presented  in
figure~\ref{fig:proton_4_10GeV_longProfile}.
The Bertini-based models give very similar results in both cases, and
are close to data. The QGSP\_BIC\_HP model predicts a reduced response
in the first calorimeter part, and a somewhat later shower maximum than observed in data.

\begin{figure}[t!]
\begin{minipage}[t]{0.48\linewidth}
\centering
\includegraphics[width=0.9\textwidth]{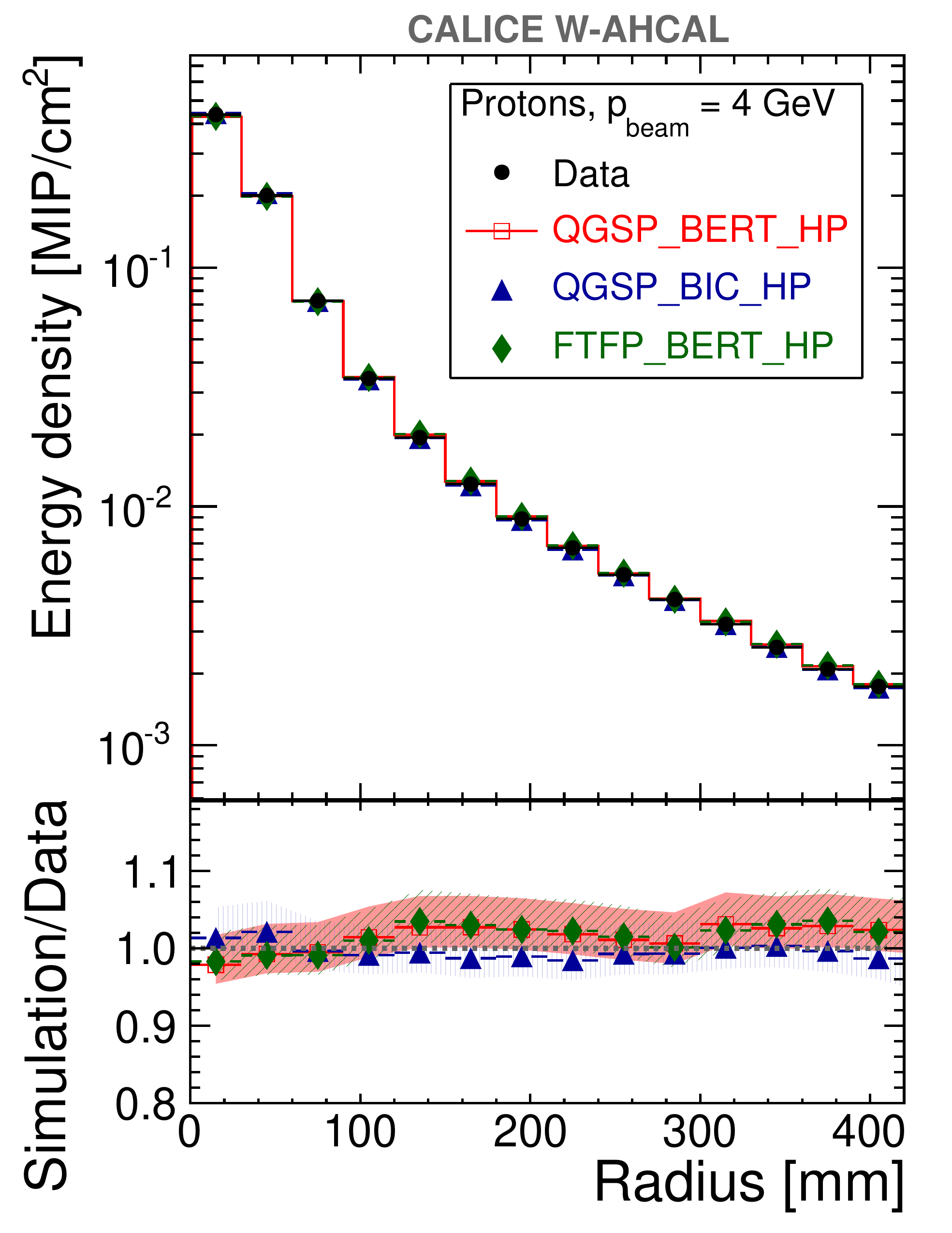}
\end{minipage}
\hspace{0.5cm}
\begin{minipage}[t]{0.48\linewidth}
\centering
\includegraphics[width=0.9\textwidth]{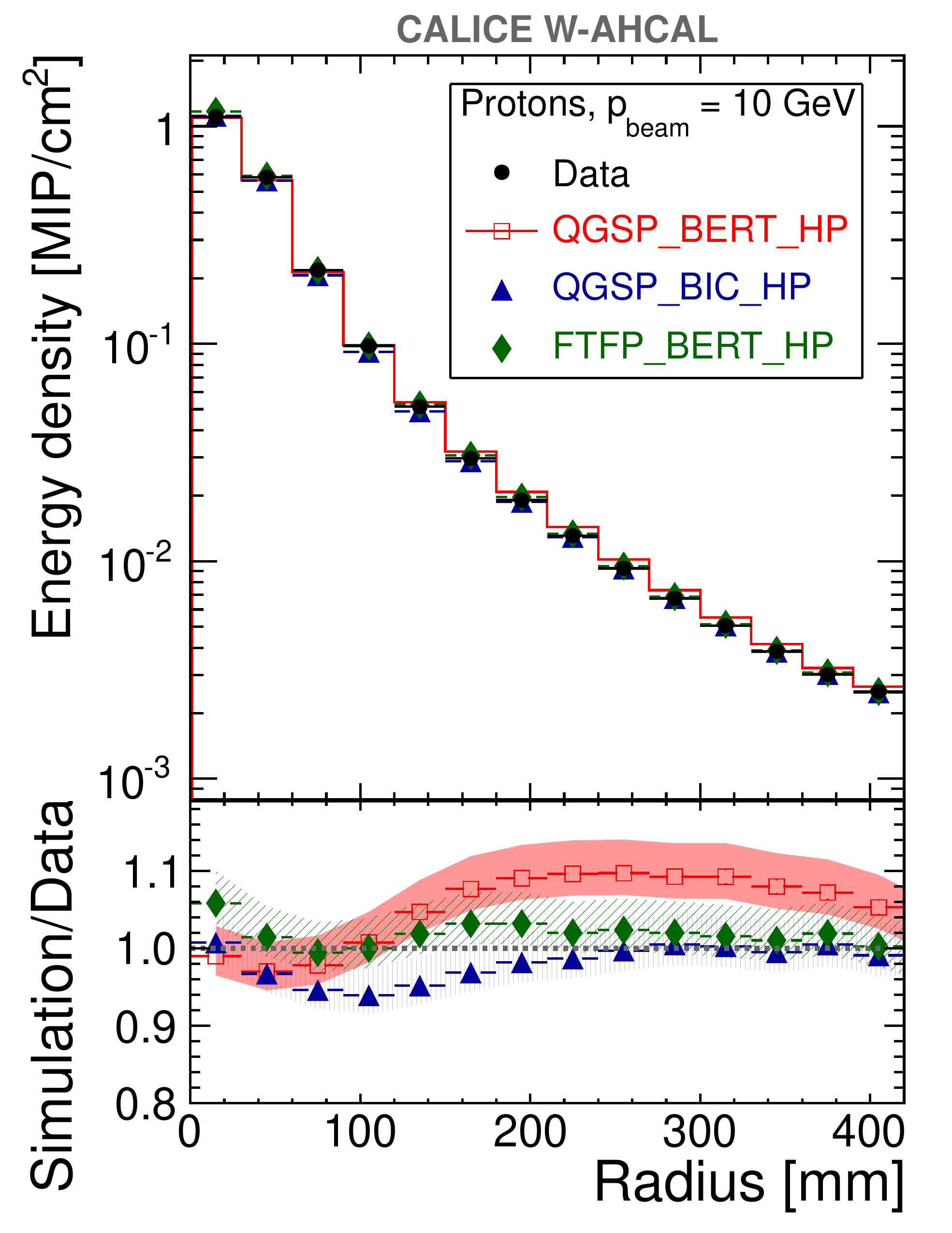}
\end{minipage}
\caption{Radial shower profiles for a proton with a beam momentum
  of 4~GeV/c (left) and of 10~GeV/c (right). Data are compared with selected
  GEANT4 physics lists. In the bottom part of the figures the ratios
  between the simulation and the data are shown.}
\label{fig:proton_4_10GeV_radialProfile}
\end{figure}

The radial shower profiles for protons with beam momentum
  of 4 and  of 10~GeV/c  are shown in
  figure~\ref{fig:proton_4_10GeV_radialProfile}. All selected physics
  lists are in agreement with the data in the 4~GeV/c case. For
  10~GeV/c, the best prediction is given by FTFP\_BERT\_HP, the
  deviations being less than 5\%. However, all physics lists show in
  this case a dependence on the shower radius.

\section{Comparison of the calorimeter response for different particle
  types}
\label{sec:response}

\begin{figure}[t!]
\begin{minipage}[t]{0.48\linewidth}
\centering
\includegraphics[width=0.9\textwidth]{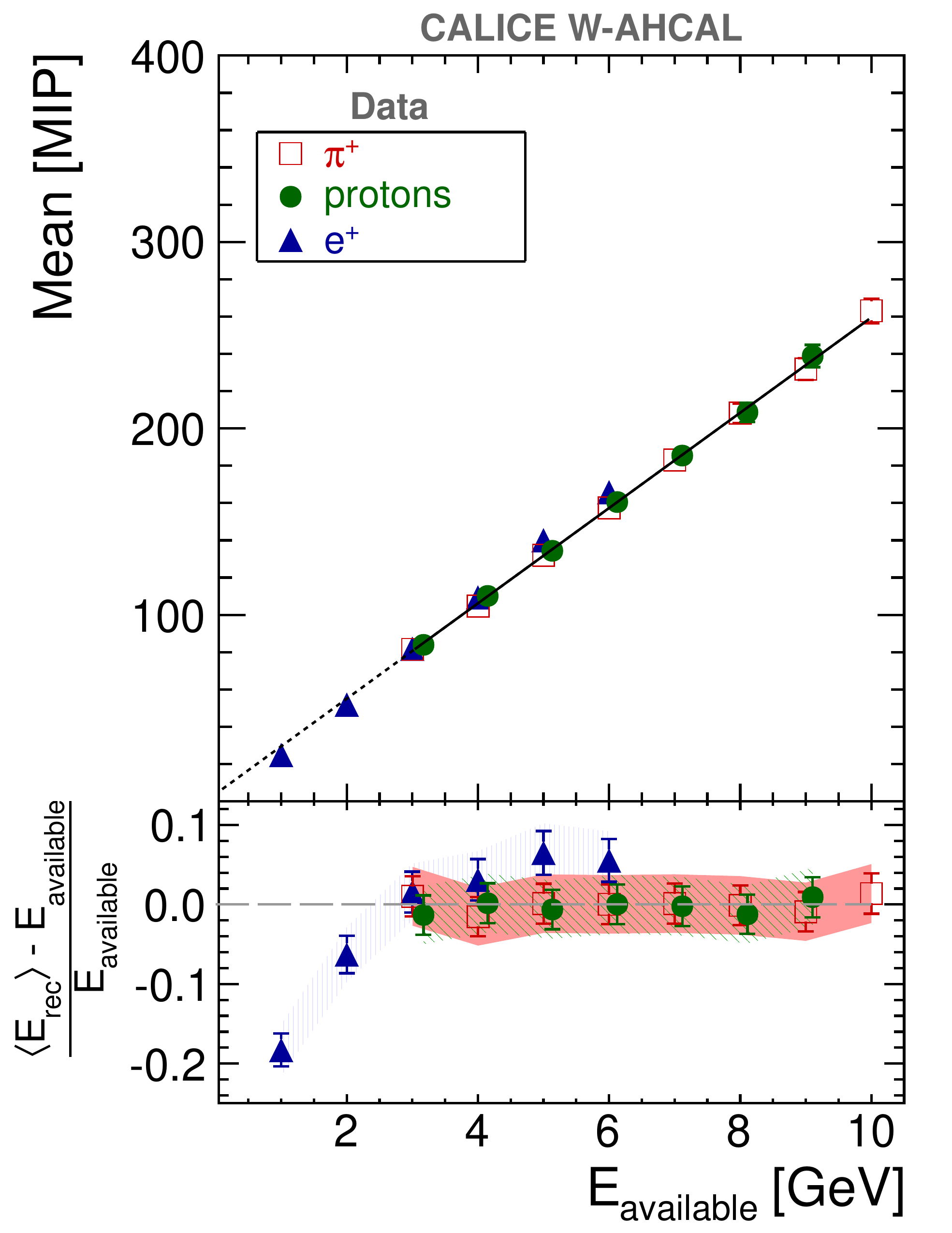}
\end{minipage}
\hspace{0.5cm}
\begin{minipage}[t]{0.48\linewidth}
\centering
\includegraphics[width=0.9\textwidth]{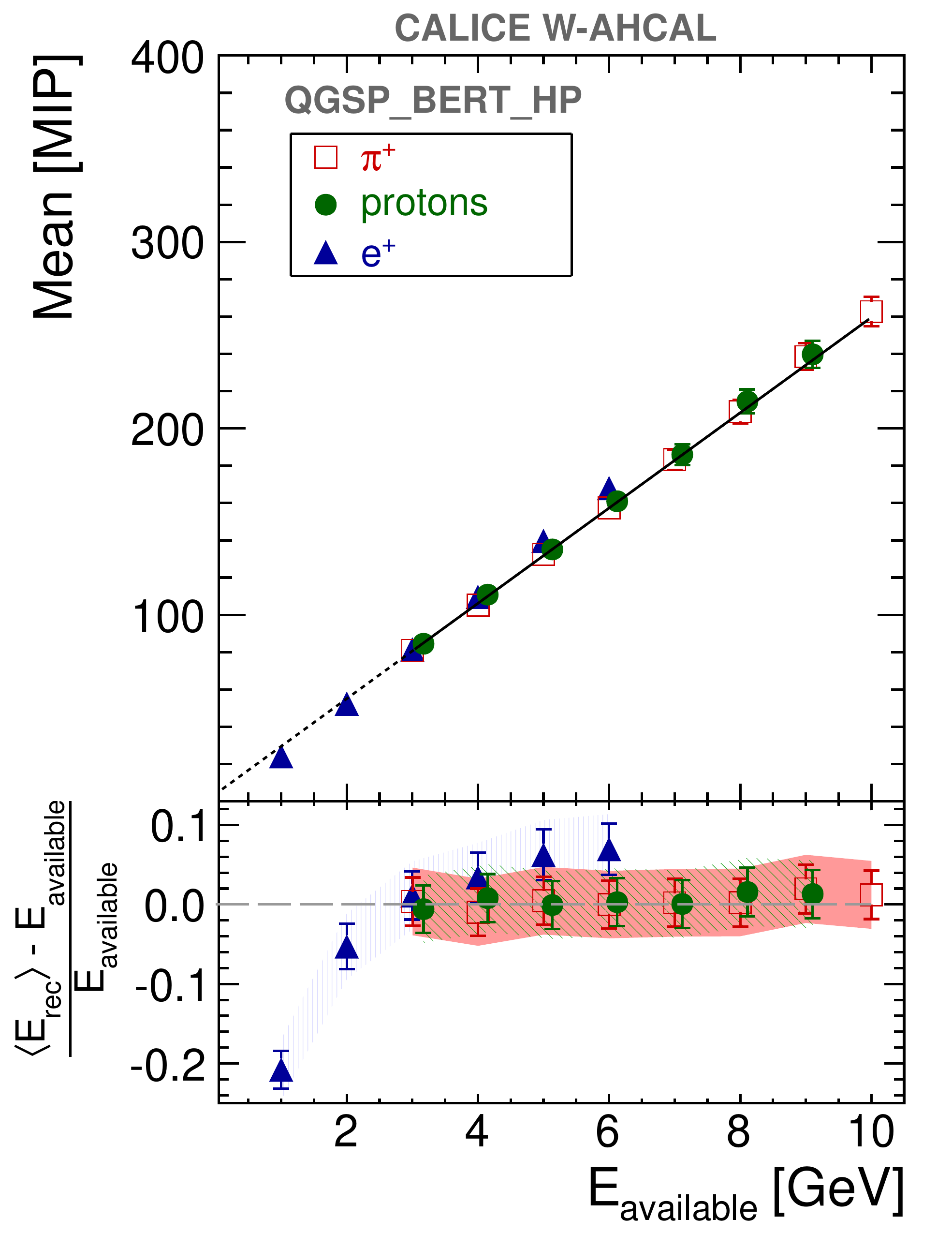}
\end{minipage}
\caption{Dependence of the mean visible energy $\langle E_{\mathrm{vis}\rangle}$ on the available
  energy for $e^+$, $\pi^+$ and protons, for data (left)
   and for QGSP\_BERT\_HP (right). 
   In the $e^+$ case,
  the mean energy is obtained from a fit, while for hadrons it is
  given by the statistical mean of the corresponding distribution. 
  The filled
  line indicates a fit of the $\pi^+$ experimental data with the
  function $\textrm{Mean} = u + v\cdot
  E_{\mathrm{available}}$. 
  The dotted line indicates the
  extrapolation of the line to zero. 
   The bottom part of the figures shows
  the residuals from this fit, where $\langle E_{\mathrm{rec}}\rangle
  \mathrm{[GeV]}=(\textrm{Mean}\;\mathrm{[MIP]} -u )/v$.
  The bands show the overall uncertainties.}
\label{fig:compensation}
\end{figure}

The calorimeter response to $\pi^+$, protons and positrons is compared
in figure~\ref{fig:compensation} for data
(left) and for simulation (right). The upper part of the figures shows
the reconstructed energy as a function of the available
energy.  The filled
  line indicates a fit of the $\pi^+$ experimental data with the
  function $\textrm{Mean} = u + v\cdot
  E_{\mathrm{available}}$, the fit parameters being given in
  table~\ref{tab:piPlus_linearityFitResults}. The corresponding fit
  parameters for protons and positrons are shown in
  figure~\ref{pLinearityData} and in
  table~\ref{tab:ePlus_linearity_mc_results}. The values for pions and
  protons are compatible, whereas the $e^+$ data show a slightly
  steeper slope. This behaviour is also predicted by the simulation.  The bottom
part of the figures shows the residuals from the linear fit of the
experimental $\pi^+$  data. 
The CALICE \mbox{W-AHCAL} response to positrons,
pions and protons is very similar from 3~GeV onwards, the differences
being smaller than $\pm 5\%$.

\begin{table}[t!]
\centering
    \caption{Fit parameters of the  dependence of the mean
    $\pi^+$ visible energy on the available energy.}
    \label{tab:piPlus_linearityFitResults}
    \begin{tabular}{ll}\toprule
      Parameter &  Value  \\\midrule
      $u$ [MIP] & $3.95 \pm 3.16$\\
      $v$ [MIP/GeV] &  $25.56 \pm 0.61$ \\
      $\chi^2$/ndf & 0.9/6 \\\bottomrule
      \end{tabular}
\end{table}

\section{Summary}
\label{sec:summary}

We presented a study of low momentum ($p_{\mathrm{beam}}\leq10$~GeV/c)
$e^{\pm}$, $\pi^{\pm}$ and proton-initiated showers in the CALICE
tungsten-scintillator analog hadron calorimeter prototype. The analysis includes measurements of the
energy resolution for the different particle types and studies of the
shower development in the longitudinal and in the transverse plane. The
energy resolution for hadrons has a stochastic term of approximately
$62\%$/$\sqrt{E\;[\mathrm{GeV}]}$
and a constant term of the order of 7\% to 11\%.
The modelling of the detector configuration and response is verified
with electrons and shows excellent agreement with the data.

The hadron
results are compared with the following {\sc geant}4 physics lists:
QGSP\_BERT\_HP, \\
FTFP\_BERT\_HP and QGSP\_BIC\_HP. The QGSP\_BERT\_HP physics list
is found to perform remarkably well for both pions and protons,
 the deviations being for most of the
studied variables within 3\% or better. In the case of protons,
 QGSP\_BIC\_HP describes both the average calorimeter
 response and the RMS of the
 visible energy distribution with reasonable accuracy.
It also agrees with data, within uncertainties,in the case of radial profiles of protons with a beam momentum of 4~GeV/c.

For available energies between 3 and 10~GeV the CALICE W-AHCAL gives
a similar response to $\pi^+$, positrons and protons.

\acknowledgments

We gratefully acknowledge the CERN technical staff: E.~Richards, I.~Krasin, 
D.~Piedigrossi, D.~Fraissard and R.~Loos for the help in the W-AHCAL
test beam. We also gratefully acknowledge the DESY and CERN
managements for their support and hospitality, and their accelerator
staff for the reliable and efficient beam operation. 
The authors would like to thank the RIMST (Zelenograd) group for their
help and sensors manufacturing.
This work was supported by the  European Commission under
the FP7 Research Infrastructures project AIDA, grant agreement no.\ 262025;
by the Bundesministerium f\"{u}r Bildung und Forschung, Germany;
by the  the DFG cluster of excellence `Origin and Structure of the Universe' of Germany; 
by the Helmholtz-Nachwuchsgruppen grant VH-NG-206;
by the BMBF, grant no.\ 05HS6VHS1;
by the Russian Ministry of Education and Science contracts 8174, 8411,
1366.2012.2, and 14.A12.31.0006;
by MICINN and CPAN, Spain;
by CRI(MST) of MOST/KOSEF in Korea;
by the US Department of Energy and the US National Science
Foundation;
by the Ministry of Education, Youth and Sports of the Czech Republic
under the projects \mbox{AV0 Z3407391}, AV0 Z10100502, LC527  and LA09042  and by the
Grant Agency of the Czech Republic under the project 202/05/0653;  
by the National Sciences and Engineering Research Council of Canada; 
and by the Science and Technology Facilities Council, UK.

\bibliographystyle{unsrt}
\bibliography{main}

\end{document}